\documentclass[11pt,tightenlines,eqsecnum,floats,aps,amsmath,amssymb,nofootinbib,prd,shownopacs,floatfix]{revtex4}
\usepackage{natbib}
\usepackage{array}
\usepackage{graphicx}
\usepackage{epstopdf}
\usepackage{latexsym}
\usepackage{amssymb}
\usepackage{amsmath}
\usepackage{color}
\usepackage{mathrsfs}
\usepackage{xparse}
\usepackage{float}
\usepackage{mathtools}
\usepackage{multirow}
\usepackage[center]{subfigure}

\usepackage{xcolor}
\usepackage{caption}
\allowdisplaybreaks

\begin{document}
\renewcommand\arraystretch{2}
 \newcommand{\bq}{\begin{equation}}
 \newcommand{\eq}{\end{equation}}
 \newcommand{\bqn}{\begin{eqnarray}}
 \newcommand{\eqn}{\end{eqnarray}}
 \newcommand{\nb}{\nonumber}
 \newcommand{\cb}{\color{black}}
    \newcommand{\cc}{\color{cyan}}
     \newcommand{\lb}{\label}
        \newcommand{\cm}{\color{magenta}}
\newcommand{\rc}{\rho^{\scriptscriptstyle{\mathrm{I}}}_c}
\newcommand{\rd}{\rho^{\scriptscriptstyle{\mathrm{II}}}_c}
\NewDocumentCommand{\evalat}{sO{\big}mm}{%
  \IfBooleanTF{#1}
   {\mleft. #3 \mright|_{#4}}
   {#3#2|_{#4}}%
}
\newcommand{\PRL}{Phys. Rev. Lett.}
\newcommand{\PL}{Phys. Lett.}
\newcommand{\PR}{Phys. Rev.}
\newcommand{\CQG}{Class. Quantum Grav.}
\newcommand{\parallelsum}{\mathbin{\!/\mkern-5mu/\!}}

\title{Primordial Gravitational Waves in Loop Quantum Cosmology: \\ Regularization Ambiguities and Initial State Dependence}
\author{Mah Noor Jamil$^1$}
\email{mjamil4@lsu.edu}
 \author{Parampreet Singh$^{1,2}$}
\email{psingh@lsu.edu}
\affiliation{$^1$ Department of Physics and Astronomy, 
Louisiana State University, Baton Rouge, LA 70803, USA}
 \affiliation{$^2$ Center for Computation and Technology,
Louisiana State University, Baton Rouge, LA 70803, USA}

\begin{abstract}
We compute the primordial tensor power spectrum in standard loop quantum
cosmology and in the modified models mLQC-I and mLQC-II, evolving the tensor
perturbations across the bounce in the hybrid and dressed metric prescriptions
with a Starobinsky potential. In mLQC-I, the emergent de Sitter contracting branch requires the use of dedicated initial states. We use  de Sitter mode function in the dressed metric
approach and a Hankel-function solution in the hybrid approach. 
For inflationary e-foldings 
$N_e = 60$ and $65$, the pivot-scale tensor-to-scalar ratio agrees with the value of standard inflation with this potential. 
The scale-dependent ratio $r(k)=\mathcal{P}_T/\mathcal{P}_S$ cancels effects
common to the two sectors and separates features that are independent of the
initial state from those that depend on it. It varies slowly in the
scale-invariant regime and oscillates around $r_\star$ at smaller $k$. The
oscillations next to the scale-invariant regime do not depend on the adiabatic
order, unlike the peaks and spikes at lower $k$. 
For a given initial state, the regularization dependence survives in the infrared, 
where long-wavelength modes retain a record of the
bounce and of the pre-bounce branch. We propagate the normalized spectra to the
CMB $B$-mode spectrum and discuss observational prospects.
\end{abstract}

\maketitle

\section{Introduction}

The big bang singularity is one of the central problems of classical general
relativity. In the standard description, the scale factor vanishes and the
curvature and energy density diverge when the cosmological evolution is
extrapolated backward in time. Inflation accounts for the origin of the
large-scale structure of the universe and for the observed near scale-invariance
of the primordial scalar perturbations. It does not remove the singular behavior
of the classical background. The past incompleteness of sufficiently expanding
spacetimes shows that inflation cannot be regarded as a complete description of
the earliest stage of cosmic evolution \cite{Borde:2001nh}. Loop quantum
cosmology (LQC) applies the quantization methods of loop quantum gravity to
homogeneous cosmological spacetimes and replaces the classical singularity with
a quantum bounce
\cite{Ashtekar:2006wn,Ashtekar:2006uz,Ashtekar:2006rx,Ashtekar:2011ni}. 
In LQC, the energy density is bounded above by a critical density of Planckian order, and
a nonsingular bounce connects the contracting branch to the expanding branch. It turns out
that all strong curvature singularities, at which geodesic evolution would break down, are
resolved in the effective spacetime description of LQC, in isotropic spacetimes with and
without spatial curvature and in anisotropic  spacetimes
\cite{Singh:2009singular,Singh:2011exotic,Singh:2012bianchi,Saini:2016ks}. This is in
contrast with the classical past incompleteness noted above.
This gives a well-defined setting in which the pre-inflationary quantum geometry
can influence the perturbations that later seed the cosmic microwave background
(CMB). Most of the modes observed in the CMB evolve in a regime where general
relativity is an excellent approximation. The largest observable scales can still
retain information about the dynamics before and during the bounce. This
possibility makes the resolution of the singularity in LQC a framework that can
be confronted with observation.

The effective spacetime description is the main tool for such phenomenological
studies. It captures the highly quantum regime near the classical singularity
with modified, nonsingular equations for the expectation values of the
cosmological variables, and it retains the characteristic holonomy corrections of
the underlying quantum theory. Extensive analytical and numerical studies show
that it describes the evolution of sharply peaked quantum states accurately,
including through the quantum bounce \cite{Ashtekar:2011ni,DienerGuptSingh2014}.
Cosmological perturbations on such effective backgrounds have been formulated
mainly in two approaches. In the dressed metric approach, the homogeneous
geometry is quantized first, and the perturbations are treated as quantum fields
propagating on an effective metric constructed from expectation values of
background operators \cite{Agullo:2012sh,Agullo:2013ai,Li:2019qzr}. In the hybrid
approach, the background and perturbative degrees of freedom are kept together as
a constrained canonical system. The homogeneous geometry is quantized with loop
methods and the perturbations as Fock degrees of freedom
\cite{Fernandez-Mendez:2013jqa,Gomar:2014faa,Gomar:2015oea,Li:2020mfi}. Both
approaches lead to mode equations of generalized harmonic-oscillator form. The
effective mass in the equation depends on how the perturbative variables and the
background quantum corrections are defined. The polymerization of the background
variables does not in general commute with the calculation of the Poisson
brackets that generate the perturbation equations. This non-commutativity
produces different effective mass functions in the hybrid and dressed metric
descriptions, although the two approaches agree when the quantum corrections are
negligible \cite{Li:2022evi,Li:2023res}. The difference is localized mainly in
the Planck regime, where the energy density is comparable to the critical
density. Once the universe enters the classical slow-roll regime, both approaches
reduce to the usual perturbation equations of general relativity. Comparing the
two approaches helps identify which features of the primordial spectrum are
robust consequences of the background dynamics and which depend on the
quantization of the perturbations.

These approaches have motivated a substantial body of work on cosmological
perturbations in LQC. It includes studies of the primordial scalar spectra and of
the choice of initial state
\cite{Agullo:2015tca,Ashtekar:2016wpi,Ashtekar:2016pqn,MartindeBlas:2016qqj} of
inflation with the Starobinsky potential \cite{Bonga:2015xna}, and of the
resulting imprints on the CMB \cite{Li:2021overview}. The temperature and
$E$-mode polarization spectra are dominated by scalar perturbations, so most of
this work concerns the scalar sector. There, some observables have not yet been
measured with sufficient precision, and others, such as the large-angle
temperature spectrum, show anomalies that may require physics beyond standard
inflation. The possible alleviation of these anomalies and the associated
non-Gaussianity have been studied in LQC
\cite{Ashtekar:2020gec,Ashtekar:2021izi,Agullo:2018dapor,Agullo:2020cvg,Agullo:2020fbw,MartinBenito:2021szh}.
Tensor spectra have received less attention and have been compared across
approaches only more recently
\cite{Bolliet:2015bka,BenitezMartinez:2016ioj,MenaMarugan:2024tensor}. A natural
question is whether the tensor sector, which probes the same quantum geometry
through an independent set of observables, carries information about the bounce
that complements the scalar sector.

The need for such an independent test is sharpened by the non-uniqueness of the
quantization of the Hamiltonian constraint in LQC. In the standard construction,
an identity valid for the spatially flat background combines the Euclidean and
Lorentzian parts of the classical Hamiltonian constraint before the connection is
polymerized. If the Lorentzian term is instead retained explicitly and
regularized independently, following Thiemann's regularization in loop quantum
gravity, two alternative prescriptions are obtained. They are called modified
loop quantum cosmology-I (mLQC-I) and modified loop quantum cosmology-II
(mLQC-II)
\cite{Yang:2009fp,Dapor:2017rwv,Assanioussi:2019iye,Li:2018fco,LiSinghWang2018}.
At the quantum level, the Hamiltonian constraint of the two modified models
yields a fourth-order difference equation instead of the second-order equation of
standard LQC \cite{Saini:2019vonNeumann}. At the effective level, the modified
Friedmann equation contains terms of higher order in the energy density than the
quadratic term of standard LQC, and the maximum energy density at the bounce
differs \cite{LiSinghWang2018}. All three models resolve the big bang
singularity, and slow-roll inflation occurs generically after the bounce in each
of them, with a probability of failure of order $10^{-5}$ or smaller
\cite{Li:2018fco,Li:2019genericness}. Their effective background dynamics differ.
Standard LQC has a symmetric bounce, and mLQC-II has a qualitatively similar
contracting and expanding structure with a different scale for the quantum
corrections. The mLQC-I model is more unusual. Its pre-bounce branch approaches
an emergent de Sitter regime with an effective cosmological constant of
Planckian magnitude \cite{LiSinghWang2018,Li:2018fco}, which is of purely quantum
gravitational origin and has no classical limit. The bounce is therefore
asymmetric: a quantum de Sitter-like contracting branch connects to an expanding
branch that approaches the classical regime and can support slow-roll inflation.
This requires initial data that place the de Sitter phase in the pre-bounce
branch \cite{LiSinghWang2018}. The Planckian cosmological constant has further
consequences. A cyclic evolution in mLQC-I is possible only if the potential is
negative and almost Planckian in magnitude \cite{Li:2022cyclic}. In the Bianchi-I
generalization of mLQC-I, the de Sitter phase damps the anisotropic shear and
still does not lead to a classical post-bounce universe
\cite{Motaharfar:2026jcj}. These differences are physical
\cite{Li:2021overview,Li:2023handbook}, since they change the time dependence of the
scale factor, the curvature, and the Hubble rate near the bounce, and so change
the effective potential felt by cosmological perturbations. The critical
densities of the three prescriptions also set the characteristic scale at which
quantum-geometric effects can enter the primordial power spectrum.

The scalar sector shows that the regularization ambiguity has consequences. The
primordial spectra of all three models are scale-invariant in the ultraviolet
regime. In the infrared and intermediate regimes, the regularization and the
perturbation approach lead to differences of at least ten percent. For mLQC-I,
the infrared amplitude is enhanced to the Planck scale in the dressed metric
approach, where it is set by the Planckian de Sitter phase of the contracting
branch, and suppressed relative to standard LQC in the hybrid approach
\cite{Li:2019qzr,Li:2020mfi,Li:2021overview}. A systematic comparison of the
three models in both approaches, with the same Starobinsky potential and the same
number of e-folds, shows that these differences carry over to the large-angle CMB
spectrum. For adiabatic initial states, mLQC-I shows the smallest deviation from
the $\Lambda$CDM prediction at large angles, and mLQC-II shows the largest and is
the most disfavored \cite{Li:2024lqc}.

These scalar results do not settle the tensor case. In contrast to the scalar sector, whose effective mass depends on the inflaton potential and on the coupling of the inflaton perturbation to the metric, the tensor effective mass depends only on the background. The two sectors nevertheless evolve on the same background. A scalar feature can therefore originate from the background, from the scalar-specific terms in the effective mass, or from the choice of initial state, and the scalar spectrum alone cannot separate these. The tensor spectrum on identical backgrounds can: a feature that reappears in the tensor spectrum is tied to the background geometry, whereas a feature that does not is specific to the scalar sector. It is pertinent to ask whether the dependence on regularization, perturbation approach, and initial state found for the scalar modes reappears in the tensor sector, and how large it is in the tensor spectrum and in the $B$-mode spectrum.

A third issue, which is decisive in mLQC-I, is the choice of the initial state.
The pre-inflationary dynamics in LQC departs radically from quasi-de Sitter
expansion, so there is no privileged vacuum of Bunch--Davies type for the
perturbations \cite{Ashtekar:2016pqn}, and the initial state must be specified
separately. The scalar spectra show that this choice matters. Their infrared and
amplified oscillatory parts depend strongly on the regularization, the
perturbation approach, and the time at which the initial conditions are imposed
\cite{Li:2019qzr,Li:2020mfi,Li:2024lqc}. The mLQC-I model requires special care,
because the quasi-de Sitter contracting branch changes the behavior of the mode
frequency before the bounce. The usual adiabatic or WKB construction is reliable
only when the frequency varies sufficiently slowly compared with the oscillation
time of the mode. Modes outside the effective horizon in the pre-bounce de Sitter regime do not satisfy this
condition \cite{Pan:2025mlqci}, and for them the initial state should not be taken to be
adiabatic. An apparent enhancement or spike from
a high-order adiabatic state in this situation may reflect the failure of the
approximation. It should not be read as a physical prediction of the model. A more
appropriate prescription for the modes of interest is a de Sitter mode function
based on the asymptotic form of the mLQC-I background. Applying the
Birrell--Davies method in the remote contracting phase, Pan, Saeed, and Wang
showed that this state is stable in the dressed metric approach, minimizes
particle creation, and diagonalizes the Hamiltonian, even though some modes are
still outside the Hubble horizon and not in their adiabatic states. They also
obtained analytic first-order scalar mode functions with the uniform asymptotic
approximation, and noted that the generalization to tensor modes is
straightforward while the hybrid approach remains to be studied
\cite{Pan:2025mlqci}. The same background branch controls the initial-state
problem in the tensor sector, so the effect of the model must be separated from
the effect of the state before any comparison with observation.

Tensor perturbations provide a natural setting for addressing all three issues.
They are the transverse-traceless perturbations of the spatial metric and do not
couple directly to the inflaton at linear order. After expansion in Fourier modes
and the introduction of the canonical variable, each mode satisfies a
second-order equation whose pump field is determined by the scale factor. The
effective mass depends only on the background geometry and on the choice of
perturbation approach. In the dressed metric approach, it is built from the Hubble rate and its time derivative on
the effective background, whereas the hybrid mass is written in terms of the energy density and pressure of the scalar field, obtained from the second-order Hamiltonian after the background variables have been polymerized. The two masses coincide in the classical limit and differ in the high-density
regime around the bounce. In mLQC-I they also differ in the contracting branch,
where the dressed metric mass approaches the de Sitter value and the hybrid mass
does not. The tensor sector also has an observational counterpart. Scalar
perturbations do not source $B$-mode polarization at linear order, so the CMB
$B$-mode spectrum is determined by the tensor modes alone. The present bound on
the tensor-to-scalar ratio is $r < 0.036$ \cite{BICEPKeckData}, and the planned
LiteBIRD mission aims at a total uncertainty $\delta r < 10^{-3}$
\cite{LiteBIRD}. The amplitude at the pivot scale is fixed by the inflationary
phase. The question for LQC is whether the bounce leaves a model-dependent
imprint at larger scales, and whether such an imprint reaches the $B$-mode
spectrum.

A detailed study of tensor perturbations in standard LQC compared the hybrid and
dressed metric approaches and found nearly identical predictions for the
ultraviolet modes \cite{MenaMarugan:2024tensor}. Analytic approximations with a
P\"oschl--Teller effective potential indicate that the infrared suppression has a
similar shape across LQC regularizations, and that the wavenumber at which it
begins scales approximately as the square root of the critical density
\cite{Kowalczyk:2025lqc}. These results suggest that the main distinction between
the three models is a shift in the scale at which quantum-geometric effects
appear, with the spectral shape largely unchanged. The existing tensor analyses do
not compare standard LQC, mLQC-I and mLQC-II systematically in both perturbation
approaches, with the same inflationary potential and background evolution. A
complete calculation with a Starobinsky potential, a consistent mLQC-I initial
state in both approaches and the corresponding $B$-mode angular spectrum has not
been carried out either. It remains open whether the differences between the
regularization prescriptions survive the bounce and the projection onto an
observable CMB quantity.

The tensor-to-scalar ratio is the quantity bounded by the data and is usually
quoted at a single pivot scale. In LQC the primordial spectra are modified at
small wavenumbers, so the ratio is a function of $k$ and the pivot value is one
point on this curve. The scale-dependent ratio
$r(k) = \mathcal{P}_T(k)/\mathcal{P}_S(k)$ is a useful diagnostic for two
reasons. Both spectra are normalized in the same way, so any effect common to the
scalar and tensor sectors cancels. What remains is the difference in how the two
sectors respond to the bounce, which originates from their different effective
masses. In addition, $r(k)$ separates the part of the small-$k$ structure that
does not depend on the initial state from the part that does, a separation that
the initial-state problem of mLQC-I makes necessary. To our knowledge, the scale
dependence of $r$ has not been studied for these models.

The goal of this manuscript is to compute the primordial tensor power spectrum in
standard LQC, mLQC-I, and mLQC-II with both the hybrid and dressed metric
approaches, and to study the tensor-to-scalar ratio and the CMB $B$-mode spectrum
that follow from it. We use the Starobinsky potential and choose the background
initial data so that the subsequent evolution produces approximately sixty
e-folds of inflation. We also check the results for sixty five e-folds, the value
used in the scalar analysis \cite{Li:2024lqc}. For each model, we derive the
effective tensor mass in both approaches, evolve the tensor modes through the
high-curvature regime, and compare the resulting spectra in the infrared,
amplified oscillatory, and scale-invariant regimes (defined in Sec.~V).

The hybrid and dressed metric masses differ mainly near the bounce, and the two
approaches agree closely in the scale-invariant regime, which includes the
observational pivot scale. The largest differences occur in the infrared, where
the modes are sensitive to the detailed pre-bounce evolution. Adiabatic initial
states of second and fourth order produce isolated spikes that come from the
choice of state. In mLQC-I the quasi-de Sitter contracting branch makes the
adiabatic prescription unreliable for superhorizon modes, so we use the de Sitter
mode function in the dressed metric approach and a Hankel-function solution in the
hybrid approach (Sec.~IV). We calculate the tensor-to-scalar ratio at the pivot
scale and as a function of $k$, and the CMB $B$-mode angular power spectrum. For
$N_e = 60$ and $65$, the pivot-scale ratio is $r_\star \simeq 4.3\times10^{-3}$
for all three models. This is close to the standard inflationary prediction and
about four times the LiteBIRD target uncertainty. For $N_e = 60$, the model
dependence of the $B$-mode spectrum is confined to the largest angular scales
($\ell \lesssim 20$), where current data cannot separate the model curves and
cosmic variance limits future measurements. These results provide a consistent
numerical framework for separating features that depend on the regularization
from features that depend on the perturbation quantization or the choice of
initial state.

The paper is organized as follows. Sec.~II reviews the effective background
dynamics of standard LQC, mLQC-I, and mLQC-II. Sec.~III derives the tensor
perturbation equations in the hybrid and dressed metric approaches and the
corresponding effective tensor masses. Sec.~IV specifies the initial conditions
for the inflaton and the background geometry, discusses the choice of initial
state with emphasis on the failure of the adiabatic expansion in the mLQC-I
contracting branch, and describes the numerical methods. Sec.~V presents the
primordial tensor power spectra for the three models and the two perturbation
approaches, computes the tensor-to-scalar ratio at the pivot scale, and studies
its dependence on $k$. Sec.~VI computes the $B$-mode angular power spectrum and
discusses the observational implications. Sec.~VII summarizes the conclusions and
describes possible extensions of this analysis. Throughout, we use units $\hbar = c =1$ and write $\kappa=8\pi G$. For numerics, the units are Planck units.

\section{Effective Background Dynamics of Loop Quantum Cosmologies}

In this section we review the effective background dynamics on which the tensor modes
propagate. We first introduce the classical phase space, the matter content and the
classical Hamiltonian constraint, and then summarize the effective constraints of
standard LQC, mLQC-I and mLQC-II. The three models differ in the treatment of the
Lorentzian part of the Hamiltonian constraint. In standard LQC it is combined with the
Euclidean part, using the identity valid for spatially flat spacetimes, before the
connection is polymerized. In mLQC-I and mLQC-II it is retained and regularized
separately, in two different ways. We concentrate on the equations of motion, the
modified Friedmann equations, the critical densities, and the classical limits. The background solutions for the three models are shown in Fig. 1. For a
more detailed discussion of the dynamics of LQC and mLQCs, see Refs. \cite{Li:2018fco,LiSinghWang2018}.

\subsection{Matter Content and Background Variables}

We consider a spatially flat, homogeneous and isotropic Friedmann-Lema\^{\i}tre-Robertson-Walker (FLRW) spacetime with a massive scalar field $\phi(t)$ that drives inflation with the Starobinsky potential, defined as:
\begin{align}
    U(\phi) = \frac{3 m^2}{4 \kappa} \left[1 - \exp\left(-\sqrt{\frac{2 \kappa}{3}}\phi\right)\right]^2,\label{eq:starobinskypotential}
\end{align}

where $m$ is the inflaton mass. The classical phase space is spanned by four variables $\{v,b,\phi,\pi_\phi\}$, where
$v=a^3$ is the volume of the fiducial cell (whose coordinate volume is set to unity),
$b$ is the variable conjugate to $v$ (on classical solutions $b=\gamma H$), and
$\pi_\phi$ is the momentum conjugate to $\phi$. The nonvanishing Poisson brackets are
$\{b,v\}=\kappa\gamma/2$ and $\{\phi,\pi_\phi\}=1$, where $\gamma\approx 0.2375$ is the
Barbero--Immirzi parameter, fixed using the black hole entropy calculation in loop
quantum gravity. The classical effective Hamiltonian constraint that establishes the dynamics of such a spatially flat FLRW universe with a scalar field is of the form:
\begin{align}
    \mathcal{H} = -\frac{3vb^2}{\kappa \gamma^2} + \frac{\pi^2_{\phi}}{2v} + vU(\phi),\label{classical-hamiltonianconstraint}
\end{align}

where $\frac{\pi^2_{\phi}}{2v} + vU(\phi) \equiv \mathcal{H}_{M}$ comprises the matter Hamiltonian. Additionally, we can define the equations for energy density $\rho$ and pressure $P$ of the scalar field:
\begin{align}
    \rho &= \frac{\pi^2_\phi}{2 v^2} + U(\phi),\label{eq:energydensity}
    \\
    P &= \frac{\pi^2_\phi}{2 v^2} - U(\phi).\label{eq:pressure}
\end{align}

We use units $\hbar=c=1$ and keep $\kappa=8\pi G$ explicit in the analytic expressions.
In the numerical evolution, all quantities are in Planck units ($G=1$). The scale factor
is normalized at the bounce, $a_B=1$.  The value $\phi_B$ of the scalar field at the
bounce is chosen separately for each model, so that all three models have the same
number of inflationary e-folds $N_e$, measured from the onset of slow-roll inflation to
its end. 

\subsection{Standard LQC}

As noted in Sec. I, the classical Hamiltonian constraint of loop quantum gravity
consists of a Euclidean and a Lorentzian term. In the spatially flat case the Lorentzian
term is proportional to the Euclidean term, so both can be combined before the loop
quantization. This gives the effective Hamiltonian constraint of standard LQC,
\begin{align}
    \mathcal{H}_{\text{LQC}} = -\frac{3v \sin^2(\lambda b)}{8 \pi G \gamma^2 \lambda^2} + \frac{\pi^2_{\phi}}{2v} + vU(\phi) \approx 0\label{eq:standardlqc-hamiltonianconstraint}
\end{align}
where $\lambda^2=4\sqrt{3}\,\pi\gamma\,\ell_{\rm Pl}^2$ is the area gap, the minimum
nonzero eigenvalue of the area operator in loop quantum gravity. This constraint
follows from the classical constraint,  by the holonomy
replacement $b^2\to\sin^2(\lambda b)/\lambda^2$. The resulting Hamilton's equations are
\begin{align}
    \dot{v} &= \frac{3v}{2\gamma \lambda} \sin(2 \lambda b)\label{eq:vdot-standardlqc},
        \\
        \dot{b} &= -\frac{3\sin^2(\lambda b)}{2 \gamma \lambda^2} - \frac{\kappa \gamma \pi^2_{\phi}}{4v^2} + \frac{\kappa \gamma}{2} U(\phi) \label{eq:bdot-standardlqc},
        \\
        \dot{\phi} &= \frac{\pi_\phi}{v},\label{eq:vdot}
        \\
        \dot{\pi}_\phi &= -v \partial_\phi U(\phi).\label{eq:pi_phi_dot}
\end{align}

From these equations, one can arrive at the modified Friedmann equation for the Hubble rate $H$ as well as its time derivative:
\begin{align}
    H^2 = \frac{\kappa}{3} \rho \left(1 - \frac{\rho}{\rho_c}\right), \hspace{4mm} \dot{H} = -\frac{\kappa}{2}(P +\rho) \left(1 - 2 \frac{\rho}{\rho_c}\right)
    \label{eq:Hsq_Hdot_LQC}
\end{align}
where $\rho_c=3/(\kappa\lambda^2\gamma^2)\approx0.41\,\rho_{\rm Pl}$ is the critical
energy density, the maximum density reached at the bounce in LQC. For $\rho\ll\rho_c$
both equations reduce to their classical counterparts:
\begin{align}
    H^2 \approx \frac{\kappa }{3}\rho, \hspace{4mm} \dot{H} \approx -\frac{\kappa}{2}(P + \rho).\label{eq:Hsq_Hdot_classical}
\end{align}

\subsection{mLQC-I}
As mentioned earlier, in mLQC-I, one can retain the Lorentzian term in the classical Hamiltonian constraint explicitly and regularize it independently from the Euclidean term, which results in the following effective Hamiltonian constraint \cite{Li:2018fco}:
\begin{align}
    \mathcal{H}_{\text{I}} = \frac{3v}{\kappa \lambda^2} \left[\sin^2(\lambda b) - \frac{(\gamma^2 + 1) \sin^2(2 \lambda b)}{4\gamma^2}\right] + \frac{\pi^2_\phi}{2v} + vU(\phi).\label{eq:mlqc-1-hamiltonianconstraint}
\end{align}
Since the matter sector of the Hamiltonian constraint does not change, the equations for $\dot{\phi}$ and $\dot{\pi}_\phi$ remain the same as for LQC. However, the equations of motion for $v$ and $b$ differ significantly:
\begin{align}
    \dot{v} &= \frac{3v \sin(2\lambda b)}{2\gamma \lambda} \left[(\gamma^2 + 1) \cos(2 \lambda b) - \gamma^2\right]\label{eq:vdot-mlqc-1},
    \\
    \dot{b} &= \frac{3v \sin^2(\lambda b)}{2 \lambda^2 \gamma} \left[\gamma^2 \sin^2(\lambda b) - \cos^2(\lambda b)\right] - \frac{\kappa \gamma \pi^2_{\phi}}{4v^2} + \frac{\kappa \gamma}{2} U(\phi).\label{eq:bdot-mlqc-1}
\end{align}

The energy density $\rho$ can be obtained from its definition in Eq. (\ref{eq:energydensity}) and then requiring that Eq. (\ref{eq:mlqc-1-hamiltonianconstraint}) must vanish, resulting in:
\begin{align}
    \rho = \frac{3}{\kappa \lambda^2} \left[- \sin^2(\lambda b) + \frac{(\gamma^2 + 1)\sin^2(2 \lambda b)}{4 \gamma^2}\right]\label{eq:rho-mlqc-1}.
\end{align}

Inverting this equation results in two distinct roots for the variable $b$:
\begin{align}
    \sin^2(\lambda b_{\pm}) = \frac{1 \pm \sqrt{1 - \frac{\rho}{\rho^{I}_c}}}{2(\gamma^2 + 1)}\label{eq:sinsqroots-mlqc-1},
\end{align}
where $\rho_c^{\rm I}=\rho_c/[4(\gamma^2+1)]\approx0.097\,\rho_{\rm Pl}$. The maximum
energy density at the bounce is therefore lower than in standard LQC by the factor
$4(1+\gamma^2)\approx4.2$. The two roots coincide at $\rho=\rho_c^{\rm I}$ and describe
the two branches of the same solution: $b_+$ governs the pre-bounce (contracting) phase
and $b_-$ the post-bounce (expanding) phase. As $\rho\to0$, $\sin^2(\lambda b_-)\to0$,
which gives the classical limit, whereas $\sin^2(\lambda b_+)\to1/(1+\gamma^2)$, so the
contracting branch has no classical limit. Because of this asymmetric behavior, one can obtain distinct expressions for $H^2$ and $\dot{H}$ in the contracting and expanding branches. In particular, in the post-bounce branch we have:
\begin{align}
    H_{\text{I}}^2 &= \frac{\kappa \rho}{3} \left(1 - \frac{\rho}{\rho^{\text{I}}_c}\right) \left[1 + \frac{\gamma^2 \rho/\rho^{I}_c}{(\gamma^2 + 1)\left(1 + \sqrt{1 - \frac{\rho}{\rho^{\text{I}}_c}}\right)^2}\right],\nonumber
    \\
    \dot{H}_{\text{I}} &= \frac{\kappa (P + \rho)}{2(1 + \gamma^2)} \left(2 \gamma^2 + 2 \frac{\rho}{\rho^{\text{I}}_c} - 3 \gamma^2 \sqrt{1 - \frac{\rho}{\rho^{\text{I}}_c}} - 1\right).\label{eq:Hsq_Hdot_mLQCI_postbounce}
\end{align}
It must be noted that as $\rho \ll \rho^{I}_c$, the above equations simplify to Eq. (\ref{eq:Hsq_Hdot_classical}), thereby recovering classical cosmological results well into the future of the expanding branch. On the other hand, the corresponding modified Friedmann equation in the contracting branch takes the following form:
\begin{align}
    H_{\text{I}_{\Lambda}}^2 &= \frac{\kappa \alpha \rho_{\Lambda}}{3} \left(1 - \frac{\rho}{\rho^{\text{I}}_c}\right) \left[1 + \left(\frac{1 - 2\gamma^2 + \sqrt{1 - \frac{\rho}{\rho^{\text{I}}_c}}}{4 \gamma^2 \left(1 + \sqrt{1 - \frac{\rho}{\rho^{\text{I}}_c}}\right)}\right)\frac{\rho}{\rho^{\text{I}}_c}\right],\nonumber
    \\
    \dot{H}_{\text{I}_{\Lambda}} &= \frac{\kappa (P + \rho)}{2(1 + \gamma^2)} \left(2 \gamma^2 + 2\frac{\rho}{\rho^{\text{I}}_c} + 3\gamma^2 \sqrt{1 - \frac{\rho}{\rho^{\text{I}}_c}} - 1\right),\label{eq:Hsq_Hdot_mLQCI_prebounce}
\end{align}
where $\alpha \equiv (1 - 5\gamma^2)/(\gamma^2 + 1)$ and $\rho_{\Lambda} \equiv 3/[\kappa \alpha \lambda^2 (1 + \gamma^2)^2]$. In the contracting branch, as $\rho \to 0$ in the remote past, $H^2$ and $\dot H$ do not
approach their classical counterparts in Eq. (\ref{eq:Hsq_Hdot_classical}). Instead they
tend to the limits $H^2\approx\kappa\alpha\rho_\Lambda/3$ and
$\dot H\approx \frac{-\kappa \alpha}{2}(P + \rho)$  which describe an emergent de Sitter
phase with an effective cosmological constant of Planckian magnitude. This phase is of
purely quantum-geometric origin and has no classical analogue. The bounce is therefore
asymmetric, unlike in standard LQC \cite{LiSinghWang2018}. Its consequences for the tensor
modes are discussed in later sections.

\subsection{mLQC-II}
In the framework of mLQC-II, the Lorentzian term is also kept explicitly, but the key difference here is that the extrinsic curvature is expressed in terms of the Ashtekar-Barbero connection, before conducting regularization and quantization of the variables. Then, the effective Hamiltonian obtained is:
\begin{align}
    \mathcal{H}_{\text{II}} = -\frac{12v}{\kappa \lambda^2 \gamma^2} \sin^2\left(\frac{\lambda b}{2}\right)\left[1 + \gamma^2 \sin^2\left(\frac{\lambda b}{2}\right)\right] + \frac{\pi^2_{\phi}}{2v} + vU(\phi)\label{eq:mlqc-2-hamiltonianconstraint}.
\end{align}

The corresponding equations of motion are given by:
\begin{align}
    \dot{v} &= \frac{3v\sin(\lambda b)}{\lambda \gamma} \left[1 + \gamma^2 - \gamma^2 \cos(\lambda b)\right]\label{eq:vdot-mlqc-2},
    \\
    \dot{b} &= - \frac{6 \sin^2\left(\frac{\lambda b}{2}\right) \left[1+ \gamma^2 \sin^2\left(\frac{\lambda b}{2}\right)\right]}{\lambda^2 \gamma} - \frac{\kappa \gamma \pi^2_{\phi}}{4v^2} + \frac{\kappa \gamma}{2} U(\phi)\label{eq:bdot-mlqc-2},
\end{align}
while equations for $\dot{\phi}$ and $\dot{\pi}_\phi$ remain the same. The modified Friedmann equation for the Hubble rate can then be expressed in terms of energy density:
\begin{align}
    H^2 = &\frac{2\kappa \rho}{3} \left(1 - \frac{\rho}{\rho^{\text{II}}_c}\right)\left[ \frac{1+4\gamma^2(1+\gamma^2)\rho/\rho^{\text{II}}_c}{1 + 2\gamma^2 \rho/\rho^{\text{II}}_c +\sqrt{1 + 4\gamma^2(1 + \gamma^2) \rho/\rho^{\text{II}}_c}}\right],\nonumber
    \\
    \dot{H} = &\frac{\kappa (P + \rho)}{2\gamma^2}\Big(2 \gamma^2 + 8\gamma^2(1 + \gamma^2) \frac{\rho}{\rho^{\text{II}}_c} - 3 \gamma^2 \sqrt{1 + 4 \gamma^2(1 + \gamma^2) \frac{\rho}{\rho^{\text{II}}_c}} + 3 \nonumber
    \\
    &- 3\sqrt{1 + 4\gamma^2(1 + \gamma^2) \frac{\rho}{\rho^{\text{II}}_c}}\Big),
    \label{eq:Hsq_Hdot_mLQCII}
\end{align}
where $\rho_c^{\rm II}=4(1+\gamma^2)\rho_c\approx1.73\,\rho_{\rm Pl}$. The maximum energy
density at the bounce is thus larger in mLQC-II than in either standard LQC or mLQC-I.
For $\rho\ll\rho_c^{\rm II}$, Eq.(\ref{eq:Hsq_Hdot_classical}) is recovered. Since
above dynamical equations in mLQC-II are single-valued functions of $\rho$,
this holds on both sides of the bounce. In contrast to mLQC-I, the contracting branch of
mLQC-II therefore also approaches the classical regime. The background solutions of the three models are illustrated in Fig. 1, which shows that the volume stays finite through the bounce and that the three models attain different values of $b$ in the contracting branch but approach the same classical behavior in the expanding branch.

\begin{figure}

    \centering

    \includegraphics[width=1.0\linewidth]{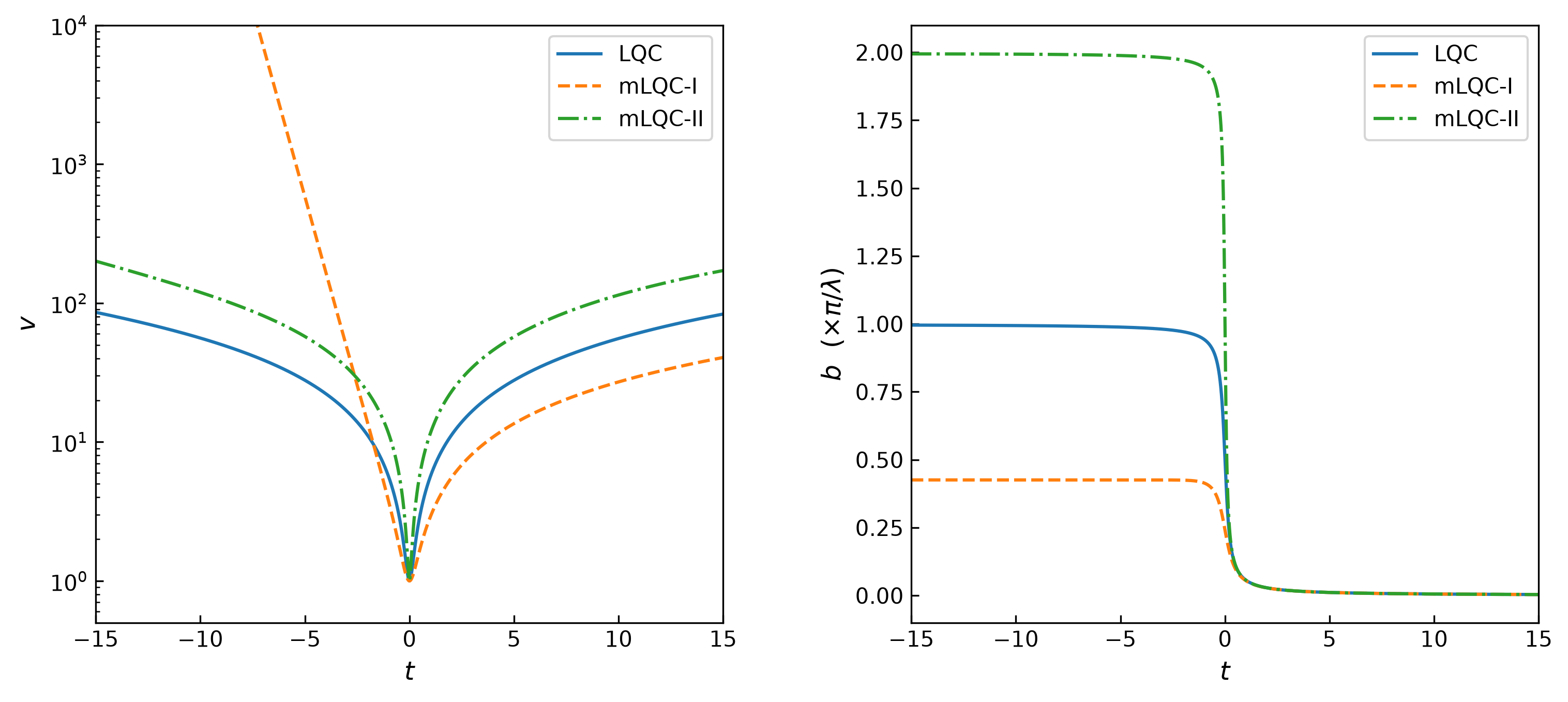}
    \captionsetup{justification=raggedright, singlelinecheck=false}
    \caption{\small On the left, we plot volume $v$ against time $t$, emphasizing their distinct behaviors near the bounce for all three models. Here, we have set the volume of the universe at the bounce to be unity, i.e. $v_B = 1$. On the right, $b$ is plotted as a multiple of $\frac{\pi}{\lambda}$ versus time $t$, showing how for all three models, $b$ attains different values in the contracting branch, but approach the same value in the expanding branch. All variables are expressed in terms of Planck units.}
\label{fig:background}
\end{figure}

\section{Tensor Perturbations}
Having reviewed the background dynamics of standard LQC, mLQC-I and mLQC-II, we now
turn to cosmological perturbations. The tensor sector is of particular interest because,
at linear order, it decouples from the scalar sector, is gauge invariant, and is the
source of primordial gravitational waves and of the CMB $B$-modes. The scalar perturbations of LQC, mLQC-I and mLQC-II with the Starobinsky potential were
studied in Ref.~\cite{Li:2024lqc}. Numerical studies of tensor perturbations on LQC 
backgrounds have so far been carried out only for standard LQC~\cite{MenaMarugan:2024tensor};
hybrid LQC predictions for the CMB were obtained in Ref.~\cite{Gomar:2017tensor}. Here we
compute the tensor modes for all three models, on the same Starobinsky backgrounds as in
Ref.~\cite{Li:2024lqc}, and ask how they evolve through the bounce and the subsequent
slow-roll inflation and what imprint this leaves on the primordial tensor spectrum and on
the $B$-mode spectrum.

In this section we introduce the tensor variables and the tensor mode equation, and then
give the effective tensor mass in the dressed metric and hybrid prescriptions. The two
prescriptions start from different treatments of the perturbations, which leads to
different effective masses, but they agree in the classical limit. Comparing them where
quantum geometry is important shows how much the perturbation prescription matters, as
distinct from the choice of regularization of the background.

\subsection{Tensor Variables and Normalization}
Cosmological perturbations around the homogeneous and isotropic FLRW universe affect both
the scalar field and the spacetime metric. Our focus is on the tensor perturbations, which
are gauge invariant at linear order and decouple from the scalar perturbations, so that the
effects of quantum geometry on gravitational waves can be isolated. Starting from the
classical Hamiltonian constraint of general relativity and truncating it at second order in
the perturbations, the tensor part of the total Hamiltonian is the sum of the background
Hamiltonian and a quadratic contribution of the tensor perturbations.

The tensor sector of the quadratic Hamiltonian describes transverse-traceless
perturbations $\gamma_{ij}$ of the spatial metric, satisfying $\gamma^i_i = 0$ and
$\partial_i \gamma^{ij} = 0$. These configuration variables can be decomposed into Fourier
modes with coefficients $\gamma^{(\epsilon)}_{k}$, where $k$ is the magnitude of the
comoving wavevector of the mode and $\epsilon$ labels the two polarizations of the
gravitational waves. Finally, we introduce a rescaled variable that is proportional to the
scale factor, defined as $h^{(\epsilon)}_{k} = a \gamma^{(\epsilon)}_{k}/2\sqrt{\kappa}$,
which puts the mode equation in the form of a generalized harmonic oscillator equation
without a friction term:
\begin{align}
    \left(h^{(\epsilon)}_{k}\right)'' + \left(k^2 + s^{(T)}\right)h^{(\epsilon)}_{k} = 0,\label{eq:MSequation-tensormodes}
\end{align}
where $s^{(T)}$ is the effective mass function for the tensor modes and primes denote
derivatives with respect to conformal time $\eta$ ($\mathrm{d}\eta = \mathrm{d}t/a$). The equation is the
same for both polarizations. The mode function is normalized by the Wronskian condition
\begin{align}
    h^{(\epsilon)}_{k}\left(h^{(\epsilon)'}_{k}\right)^{\star} - \left(h^{(\epsilon)}_{k}\right)^{\star} h^{(\epsilon)'}_{k} = i,\label{eq:wronskiancondition}
\end{align}
where the asterisk denotes the complex conjugate. The perturbation approach enters the mode
equation only through the form of the effective tensor mass $s^{(T)}$.

\subsection{Dressed-Metric Prescription}
The dressed metric approach is based on the Hamiltonian formulation of cosmological
perturbation theory in general relativity introduced by Langlois, which uses the spatially
flat gauge \cite{Langlois:1994}. In this approach the cosmological perturbations are Fock
quantized, while the background geometry is quantized with loop quantization techniques
(standard LQC or its modifications). The quantum fluctuations then propagate on a dressed
metric, which is constructed from expectation values of background operators and, in the
effective description used here, is evaluated on the effective dynamics. In this
prescription the mode equation of the tensor perturbations has the following effective mass:
\begin{align}
    s^{(T)}_{\text{DM}} = -\frac{a''}{a} = -a^2(2H^2 + \dot{H}).\label{eq:effmass-tensormodes-dressedmetric}
\end{align}

As discussed in Sec.~II, the regularization ambiguities lead to
different background dynamics. Since the linear perturbations propagate on the effective
spacetime of the corresponding model, as long as its effective dynamics are valid, the
background quantities in the effective mass must be taken from that model. Each model has
its own $H^2$ and $\dot{H}$ as functions of the energy density and pressure. For LQC and
mLQC-II, they are given by Eqs.~(\ref{eq:Hsq_Hdot_LQC}) and (\ref{eq:Hsq_Hdot_mLQCII}),
respectively. For mLQC-I they are given by Eq.~(\ref{eq:Hsq_Hdot_mLQCI_postbounce}) in the
expanding branch and by Eq.~(\ref{eq:Hsq_Hdot_mLQCI_prebounce}) in the contracting
branch. The observable modes exit the Hubble radius in the expanding branch, and the
primordial spectra are evaluated for these modes. However, the modes are initialized in the
contracting branch and traverse the bounce, so both sets
of expressions enter the evolution in mLQC-I. Each model therefore has its own effective
mass function, and the differences are expected to be largest near the bounce, where the
quantum geometry effects are strongest.

\subsection{Hybrid Prescription}

The hybrid approach builds on the gauge-invariant treatment of Halliwell and Hawking
\cite{Hawking:1984}, in which the background spacetime and the cosmological perturbations
are described together as a constrained canonical system, and different quantum
representations can be used for the different degrees of freedom. This leads to effective
equations of motion for the tensor perturbations, which can be combined into the usual
tensor mode equation with the effective mass
\begin{align}
    s^{(T)}_{\text{H}} = -\frac{\kappa}{6} a^2 \left(\rho - 3P\right),\label{effmassfunc_tensormodes_hybrid}
\end{align}
where $\rho$ and $P$ are the energy density and pressure of the scalar field in the
corresponding background model \cite{MenaMarugan:2024tensor}. As discussed in
Sec.~II, the regularization determines the background dynamics, and
hence $\rho(t)$ and $P(t)$, especially near the bounce. Each model therefore has its own
hybrid effective mass. The dressed metric mass of
Eq.~(\ref{eq:effmass-tensormodes-dressedmetric}) is built from the geometric quantities
$H^2$ and $\dot{H}$, whereas the hybrid mass is written in terms of the matter quantities
$\rho$ and $P$. In both cases the mass is fixed by the background solution alone, and it
does not involve the inflaton perturbation or its coupling to the metric. The two forms
coincide when the classical Friedmann equations hold, but not where the modified Friedmann
equations apply. This difference matters for the choice of initial state, in particular in
the contracting branch of mLQC-I.

We now compare the effective masses of the two prescriptions for each model, first near
the bounce and then during slow-roll inflation. In particular, we study how the quantum
geometry effects near the bounce make LQC, mLQC-I and mLQC-II differ in both approaches,
and in which regimes the effective masses converge during slow-roll inflation, after the
energy density has dropped far below its critical value. This allows us to separate the
effect of the perturbation prescription from the effect of the regularization of the
Hamiltonian constraint.

\subsection{Comparison of Effective Masses}

\begin{figure}
    \includegraphics[width=1.0\linewidth]{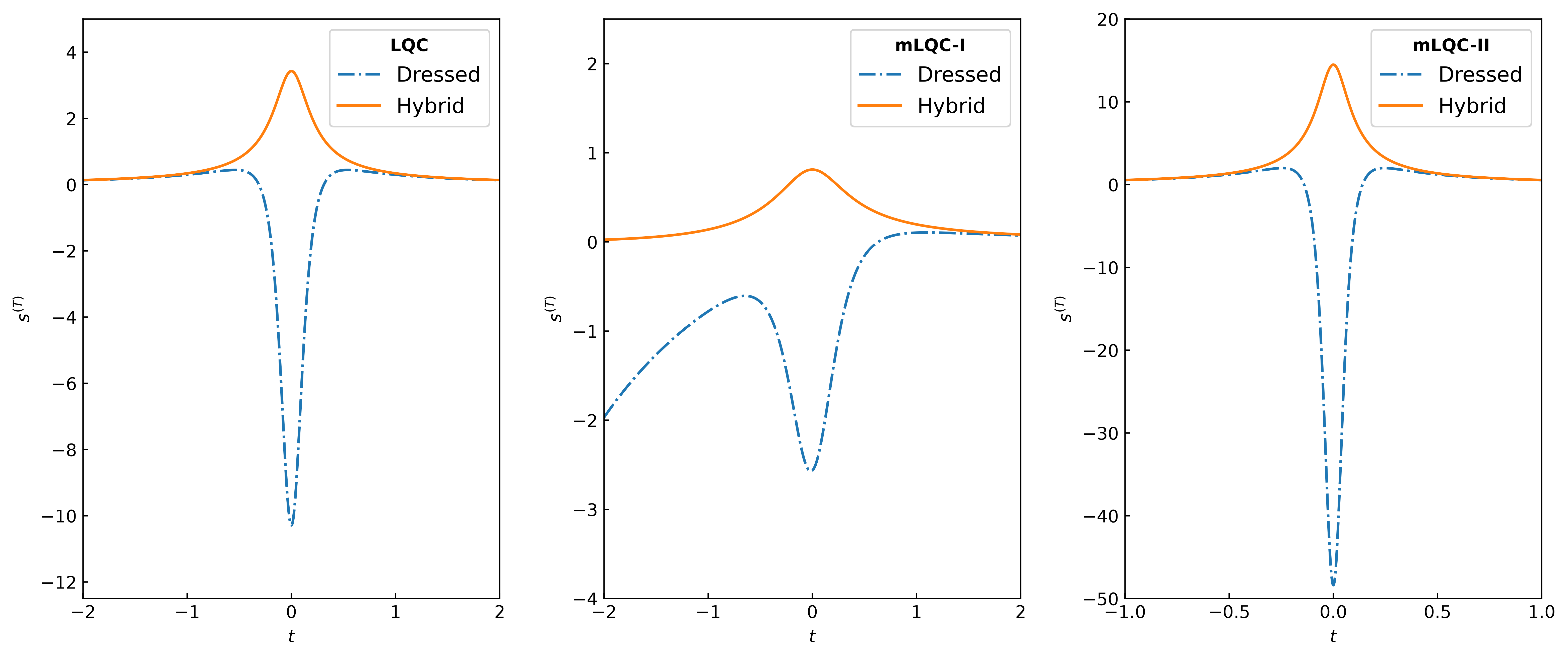}
    \captionsetup{justification=raggedright, singlelinecheck=false}
    \caption{\small Hybrid and dressed metric effective tensor masses near the bounce for LQC, mLQC-I and mLQC-II. All variables are expressed in Planck units, with $a_B=1$.}
    \label{FIG:hybridvsdressedmetric_effmassfunc}
\end{figure}

The effects of the regularization and of the treatment of the perturbations become
apparent when the effective masses are followed in time. We first compare the effective
mass functions of the dressed metric and hybrid approaches for each model in
Fig.~\ref{FIG:hybridvsdressedmetric_effmassfunc}. All three models share the feature that
$s^{(T)}_{\text{DM}}$ is negative near the bounce, with the largest magnitude for mLQC-II
and the smallest for mLQC-I. This ordering follows the critical densities of the three
models, $\rho^{\text{II}}_c > \rho_{c} > \rho^{\text{I}}_c$, which set the scale of the
quantum geometry effects at the bounce. Far from the bounce, $s^{(T)}_{\text{DM}}$ and
$s^{(T)}_{\text{H}}$ approach the same value in both the contracting and the expanding
branch for LQC and mLQC-II. For mLQC-I they agree only in the expanding branch. In the
contracting branch, $H_{\text{I}_\Lambda}^2$ and $\dot{H}_{\text{I}_\Lambda}$ tend to the
de Sitter limit instead of the classical expressions,  so that $s^{(T)}_{\text{DM}}$ tends to the de Sitter value,
while $s^{(T)}_{\text{H}}$, which is written in terms of $\rho$ and $P$, does not.

\begin{figure}
    \includegraphics[width=0.6\linewidth]{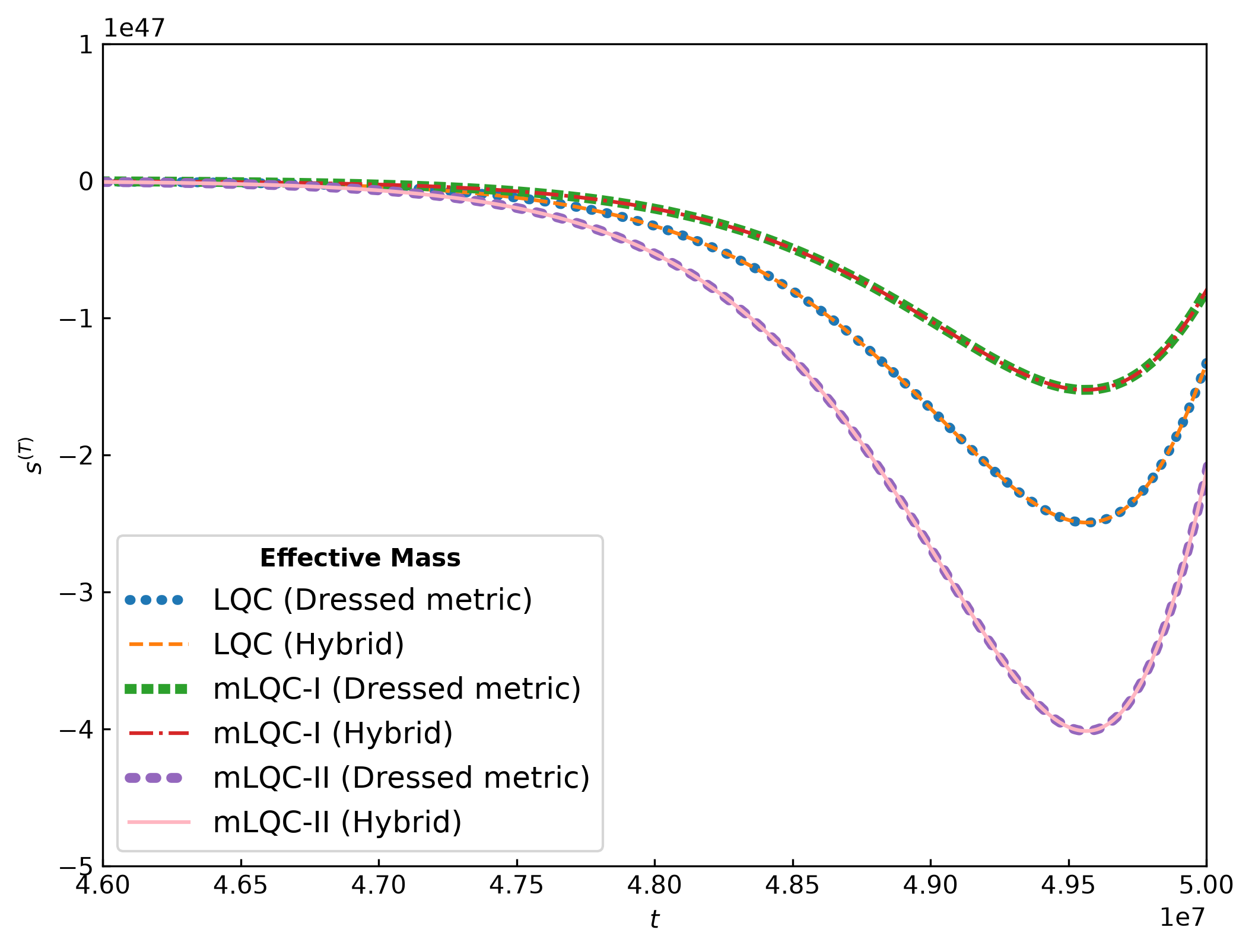}
    \captionsetup{justification=raggedright, singlelinecheck=false}
    \caption{\small Comparison of the effective tensor masses of the dressed metric and hybrid approaches for each model at the end of the inflationary period, plotted in the expanding branch. All variables are expressed in Planck units, with $a_B=1$.}
    \label{FIG:masses_relativecomparison}
\end{figure}

Fig.~\ref{FIG:masses_relativecomparison} shows the effective masses during inflation. For
all three models $s^{(T)}_{\text{DM}}$ and $s^{(T)}_{\text{H}}$ approach each other as
inflation proceeds, because $\rho \ll \rho_c$ during slow-roll inflation. In this classical
regime the tensor equation takes the form of general relativity,
\begin{align}
    \left(h^{(\epsilon)}_{k}\right)'' + \left(k^2 - \frac{a''}{a}\right)h^{(\epsilon)}_k = 0,\label{eq:MSeq-tensormodes-classical}
\end{align}
where $\frac{a''}{a} = \frac{\kappa}{6}a^2 (\rho - 3P)$ in general relativity, so that
Eqs.~(\ref{eq:effmass-tensormodes-dressedmetric}) and (\ref{effmassfunc_tensormodes_hybrid})
coincide in the classical limit for all three models. Therefore, all three models agree during most of
inflation. But near the end of inflation the effective masses separate, with
$s^{(T)}$ of mLQC-II becoming the most negative and that of mLQC-I the least negative. This
separation does not indicate different local dynamics. During slow roll all three models
have the same Hubble rate, $H\simeq m/2$, and $s^{(T)}\propto a^2$, with $a_B=1$ imposed at
the bounce in each model. At a given time the scale factors therefore differ by constant
factors, set by the expansion accumulated between the bounce and the onset of slow roll
 and by the different onset
times of slow roll. The slight increase of $s^{(T)}$ at the end of inflation occurs when the
inflaton leaves the plateau, so that $\rho-3P$ decreases, which outweighs the growth of
$a$. The regularization therefore affects the effective mass beyond the bounce, because the
pre-inflationary expansion history sets the normalization of the scale factor during
inflation. This is a normalization effect, and not a difference in the inflationary
dynamics itself.

\section{Background Initial Data, Initial States, \\ and Numerical Methods}

In this section we specify the background initial data, the initial quantum states of the
tensor modes for each model and perturbation approach, and the numerical methods used to
evolve the modes and to compute the tensor power spectrum. The background dynamics must first
be solved numerically, since the background quantities enter the tensor mode equation
(\ref{eq:MSequation-tensormodes}). We impose initial data at the bounce, which takes place
at $t = 0$. The scale factor at the bounce is set to unity, $a_B=1$, so that $v_B = 1$. The
value of $b$ at the bounce is fixed by requiring that the energy density is maximal there
in each model (Table~\ref{TAB:phi_b_vals}). The momentum of the scalar field at the bounce,
$\pi_{\phi_B}$, then follows from the requirement that the effective Hamiltonian constraint
of each model vanishes at the bounce. This gives
$\pi_{\phi_B} = \sqrt{2 \tilde{\rho}_c - 2 U(\phi_B)}$, where $\tilde{\rho}_c = \rho_c$ in
LQC, $\tilde{\rho}_c = \rho^{\text{I}}_c$ in mLQC-I and $\tilde{\rho}_c = \rho^{\text{II}}_c$
in mLQC-II. This leaves two free parameters, the inflaton mass $m$ and the field value at
the bounce, $\phi_B$, which are chosen so that a given number of inflationary e-folds is
obtained. The initial states of the tensor modes are all set in the contracting branch.

Since our goal is to compare the effects of the regularization and of the perturbation
approach, the comparison must be performed with identical background conventions. The same
inflaton mass, $m = 2.7 \times 10^{-6}$, is used for all three models, and $\phi_B$ is chosen
separately for each model so that the same number of e-folds of inflation is obtained
(Table~\ref{TAB:phi_b_vals}). 
 The normalization of the scalar spectrum, which fixes the
pivot scale and the amplitude of the tensor spectrum is described in Sec. V. The number of inflationary e-folds $N_e$ is defined as:
\begin{align}
    N_e=\ln\left[\frac{a(t_\text{end})}{a(t_\text{slow-roll})}\right],\label{eq:numofefoldings}
\end{align}
where $a(t_{\text{slow-roll}})$ and $a(t_{\text{end}})$ are the values of the scale factor
at the start and the end of slow-roll inflation, respectively. The start of slow-roll
inflation is identified by requiring that the equation of state is $w \approx -1$ and that
the Hubble slow-roll parameters $\varepsilon \equiv -\frac{\dot{H}}{H^2}$ and
$|\delta| \equiv \left|\frac{\ddot{\phi}}{H\dot{\phi}}\right|$ are both much less than unity
in all three models. The end of inflation is the time at which the equation of state rises
above $-1/3$, so that the expansion stops accelerating. The values of $\phi_B$ that were
selected are given in Table~\ref{TAB:phi_b_vals}.

\begin{table}[t]
%\label{tab:bounce_data}
\begin{ruledtabular}
\begin{tabular}{lcccc}
Model & $\tilde\rho_c$ & $\lambda b_B$ & $\phi_B\ (N_e=60)$ & $\phi_B\ (N_e=65)$ \\
\hline
LQC     & $\rho_c\simeq0.409$              & $\pi/2$                                              & $-1.43347$ & $-1.41550$ \\
mLQC-I  & $\rho_c^{\rm I}\simeq0.0969$     & $\arcsin\sqrt{1/(2+2\gamma^2)}\simeq0.759$           & $-1.31400$ & $-1.29600$ \\
mLQC-II & $\rho_c^{\rm II}\simeq1.73$      & $\pi$                                                & $-1.54190$ & $-1.52400$ \\
\end{tabular}
\captionsetup{justification=raggedright, singlelinecheck=false}
\caption{\small Initial data at the bounce ($t=0$) for the three models, in Planck units, with
$a_B=v_B=1$ and inflaton mass $m=2.7\times10^{-6}$ for all models. The maximum density
$\tilde\rho_c$ and the value $\lambda b_B$ at which it is reached follow from the model.
The field value $\phi_B$ is chosen separately for each model so that the number of
e-folds of slow-roll inflation, is $N_e=60$ or $65$. The momentum
$\pi_{\phi_B}=\sqrt{2[\tilde\rho_c-U(\phi_B)]}$ then follows from the Hamiltonian constraint.}
\label{TAB:phi_b_vals}
\end{ruledtabular}
\end{table}

\begin{figure}
    \includegraphics[width=1.0\linewidth]{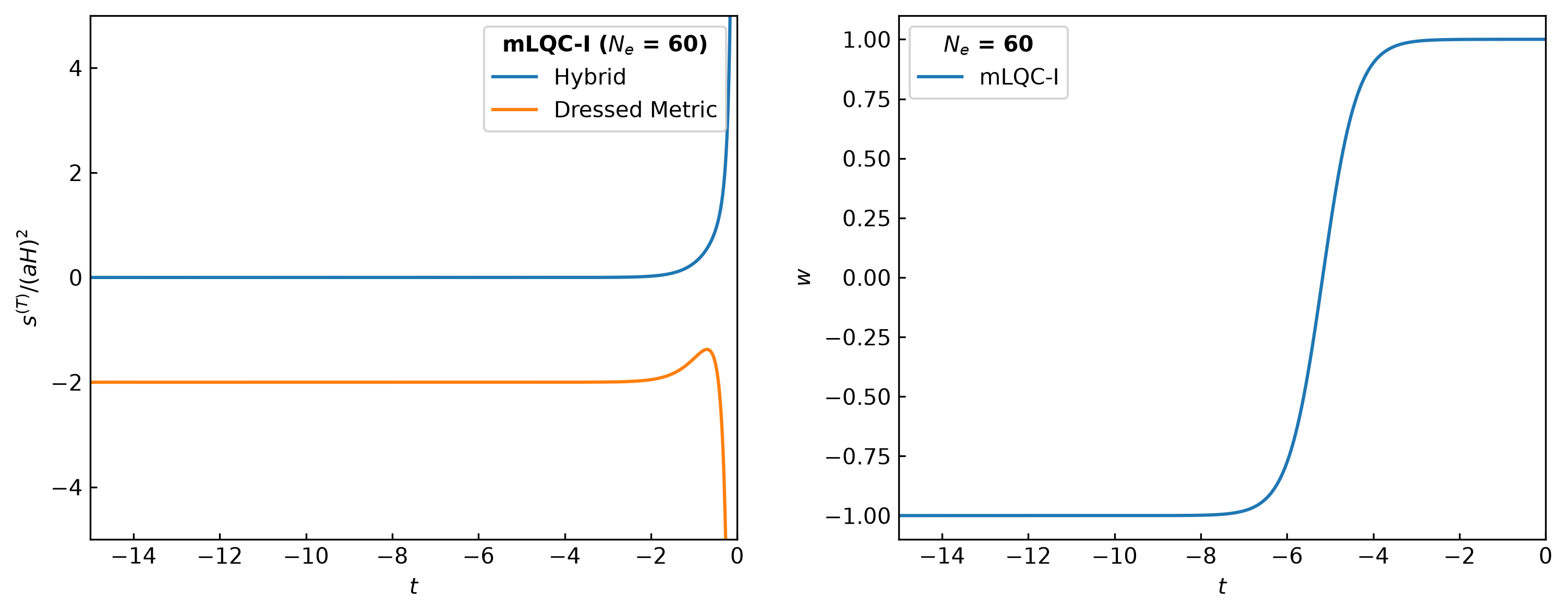}
    \captionsetup{justification=raggedright, singlelinecheck=false}
    \caption{\small Left: $s^{(T)}/(aH)^2$ versus time for the hybrid and dressed metric approaches in
the contracting branch of mLQC-I with $N_e=60$, compared with the de Sitter value
$s^{(\mathrm{dS})}/(aH)^2=-2$. Right: the equation of state $w$ of mLQC-I versus time in the
contracting branch.}
    \label{FIG:s_over_aHsq_eqofstate_mLQCI_60efolds}
\end{figure}

Once the conditions at the bounce are set, the background equations are integrated from the
bounce back to $t_{\rm start}$ in the contracting branch, and forward to the end of
inflation, using the \textsc{solve}$\_$\textsc{ivp} module from \textsc{python}'s
\textsc{scipy} package. The RK45 method is used with absolute and relative tolerances set
at $10^{-13}$. After the background dynamics are solved, the time at which each initial
state is set must be chosen with care. The initial states are set in the contracting branch
of each model, sufficiently far from the bounce so that quantum geometry effects on the
effective mass are negligible and the chosen state is a good approximation to the vacuum.
For LQC and mLQC-II we choose $t_\text{start} = -1\times 10^{6}$ (Planck units). At this time the universe in both models has approached classical general relativity,
and the effective masses of Eqs.~(\ref{eq:effmass-tensormodes-dressedmetric}) and
(\ref{effmassfunc_tensormodes_hybrid}) reduce to the same classical expression. For
mLQC-I, $t_\text{start}$ must instead be chosen deep in the contracting branch, where the
background is close to de Sitter. The initial states used there (i.e. Eqs.~(\ref{eq:deSittersolution})
and (\ref{eq:exactsol_tensormodes}), which are defined later), require that the ratio $s^{(T)}/(aH)^2$ should be constant; in
the dressed metric approach the mass is built from the geometric quantities $H^2$ and
$\dot{H}$, which take their de Sitter forms only if $H$ is constant; on the other hand, in the hybrid approach
it is built from the matter quantities $\rho$ and $P$, which do so only if $w\simeq-1$. In that regard, Fig. \ref{FIG:s_over_aHsq_eqofstate_mLQCI_60efolds} suggests that $t_\text{start} = -12$ is an appropriate choice for mLQC-I under both dressed metric and hybrid approaches.

For LQC and mLQC-II, the initial states are chosen to be the adiabatic states, which are
WKB solutions of the tensor mode equation Eq.~(\ref{eq:MSequation-tensormodes}):
\begin{align}
    h^{(\epsilon)}_k = \frac{1}{\sqrt{2 W_k}} e^{-i\int^{\eta} W_k(\bar{\eta}) d\bar{\eta}}.\label{eq:WKBsolution-general}
\end{align}
Substituting this into Eq.~(\ref{eq:MSequation-tensormodes}) gives an equation for $W_k$,
\begin{align}
    W_k^2 = k^2 + s^{(T)} - \frac{1}{2}\left[\frac{W_k''}{W_k} - \frac{3}{2}\left(\frac{W_k'}{W_k}\right)^2\right],\label{eq:Wequation}
\end{align}
which is solved order by order in the adiabatic expansion. The solutions at the zeroth,
second and fourth orders are
\begin{align}
    W^{(0)}_k = k, \hspace{4mm}W^{(2)}_{k} = \sqrt{k^2 + s^{(T)}},\hspace{4mm}W^{(4)}_{k} = \frac{\sqrt{f(s^{(T)},k)}}{4\left|k^2 + s^{(T)}\right|},\label{eq:zerothsecondfourthordersolutions}
\end{align}
where $f(s,k) = 5s'^2 + 16 k^4(k^2 + 3s) + 16 s^2 (3k^2 + s) - 4s''(k^2 + s)$, $s \equiv s^{(T)}$,
and primes denote derivatives with respect to conformal time.
These adiabatic states are well suited to LQC and mLQC-II in both the dressed metric and the
hybrid approach, because at the chosen $t_\text{start}$ the energy density is far below its
critical value, so the effective masses of the two approaches coincide.

The same cannot be said for mLQC-I. There the effective masses of the two
approaches differ in the contracting branch (Fig.~\ref{FIG:hybridvsdressedmetric_effmassfunc}).
In addition, an adiabatic state is reliable only when the mode frequency varies slowly
compared with the oscillation time of the mode. In the quasi-de Sitter contracting branch of mLQC-I, modes that are outside the Hubble
radius at $t_{\rm start}$ do not satisfy this condition \cite{Pan:2025mlqci}. For the
dressed metric mass $s\simeq-2/\eta^2$, the second order state $\sqrt{k^2+s}$ is not even
real for $|k\eta|<\sqrt{2}$. Since the effective masses differ, each perturbative approach needs its own
initial state. The left panel of Fig.~\ref{FIG:s_over_aHsq_eqofstate_mLQCI_60efolds} shows the
ratio $s^{(T)}/(aH)^2$ in the contracting branch of mLQC-I for both approaches, computed from
Eqs.~(\ref{eq:effmass-tensormodes-dressedmetric}) and (\ref{effmassfunc_tensormodes_hybrid}).
It can be compared with the constant value $s_{\text{dS}}/(aH)^2 = -2$ of the effective mass
of a de Sitter universe. The dressed metric mass matches this value best, so the de Sitter
initial state is the appropriate choice for the dressed metric approach:
\begin{align}
    h^{(\epsilon)}_{k} = \alpha_k \frac{e^{-ik\eta}}{\sqrt{2k}} \left(1 - \frac{i}{k \eta}\right) + \beta_k \frac{e^{ik\eta}}{\sqrt{2k}} \left(1 + \frac{i}{k\eta}\right),\label{eq:deSittersolution}
\end{align}
where we set $\beta_k = 0$ and $\alpha_k = 1$ in our simulations
\cite{Li:2024lqc}. The stability of this state in the dressed metric approach was shown in
Ref.~\cite{Pan:2025mlqci}. Because the de Sitter mode function is an exact
solution of the mode equation in the asymptotic de Sitter regime, and not a WKB
approximation imposed at a chosen initial time, the resulting spectrum is expected to be
insensitive to the precise value of $t_{\mathrm{start}}$ as long as $t_{\mathrm{start}}$ lies
in this regime, in contrast with the adiabatic initial states.

This state is not appropriate for the hybrid approach, because the hybrid
effective mass differs too much from the effective de Sitter mass
(Fig.~\ref{FIG:s_over_aHsq_eqofstate_mLQCI_60efolds}).
We therefore use instead the solution of the tensor mode equation
Eq.~(\ref{eq:MSequation-tensormodes}) that is exact when the ratio $s^{(T)}_{\text{H}}/(aH)^2$
is constant and $aH|\eta|$ is constant, expressed in terms of the Hankel function of the
second kind:
\begin{align}
    h_k = \frac{\sqrt{\pi |\eta|}}{2} H^{(2)}_{\mu}(k |\eta|), \label{eq:exactsol_tensormodes}
\end{align}
where
\begin{align}
\mu = \left(-\frac{s^{(T)}_{\text{H}}}{(aH)^2}\Bigg|_{t_\text{start}} + ~\frac{1}{4}\right)^{1/2}.
\end{align}
This is a more appropriate choice for mLQC-I in the hybrid approach, because it accounts for
the distinct behavior of the hybrid effective mass. As seen in
Fig.~\ref{FIG:s_over_aHsq_eqofstate_mLQCI_60efolds}, the ratio $s^{(T)}_{\text{H}}/(aH)^2$ is
close to zero and, numerically, varies by only about $10^{-4}$ in relative terms for
$t \lesssim t_{\rm start}$. The mode equation then reduces to a Bessel equation, and
Eq.~(\ref{eq:exactsol_tensormodes}) is its exact solution for that mass profile. It therefore
does not rely on a truncated adiabatic expansion whose form changes with the order. It is
exact only to the extent that this approximation holds, so some dependence on
$t_{\rm start}$ remains.

With the initial states specified for each model and perturbation approach, we compute the
primordial tensor power spectrum by feeding the background quantities and the initial states
into the tensor mode equation. We use the \textsc{pyoscode} \cite{Agocs2020} package, which is
designed to solve second-order differential equations with highly oscillatory solutions. It
uses a WKB approximation when the solution oscillates rapidly, which allows it to skip many
oscillation cycles without computing every point, and it uses an RK45 method when the
solution is non-oscillatory, with the error controlled by higher order corrections. The mode functions are obtained for wavenumbers $k \in (10^{-5},5\times 10^{4})$, in the
units in which $a_B=1$. The integration runs from $t_{\rm start}$ in the contracting branch
to the end of inflation. We then compute the primordial tensor power spectrum for the modes
that exit the Hubble radius during inflation. Since the amplitude of these modes is constant
on superhorizon scales during inflation, its value at horizon exit is the same as at the end
of inflation.

\section{Primordial Tensor Spectra}
In this section, we study the primordial tensor power spectrum of the loop quantum cosmological models under both the hybrid and dressed metric approaches. This involves first solving the tensor mode equation using the initial conditions and initial quantum states discussed in the previous sections, and then computing the tensor power spectrum for each regularization model and perturbative approach. From the figures that follow in this section, we will show that the primordial tensor power spectrum in LQC and its modified models all have three distinct regimes: the infrared regime $k \in (10^{-5},10^{-4})$, the amplified oscillatory regime $k \in (10^{-4},1)$ and the scale-invariant regime $k \in (1,5\times 10^{4})$ \cite{Li:2024lqc}. We refer to the part of the amplified oscillatory regime adjacent to the scale-invariant regime as its upper part. We compare the tensor power spectra obtained for LQC and mLQC-II first, with a focus on comparing each order of initial state one by one. In doing so, we aim to emphasize the differences in amplitudes and oscillatory behaviors in different regimes. Then, we focus on comparing the tensor power spectra obtained for mLQC-I in the hybrid and dressed metric approaches, specifically because the two perturbative approaches use distinct initial states, as discussed before. Finally, the tensor-to-scalar ratio is computed for each model under both approaches, and its dependence on $k$ is studied.

\subsection{Power-Spectrum Normalization and Inflationary Observables}
The primordial tensor power spectrum is defined as:
\begin{align}
    \mathcal{P}_{T}(k) = 2 \mathcal{P}^{(\epsilon)}_{T}(k) = \frac{4 \kappa k^3}{\pi^2}\frac{|h^{(\epsilon)}_{k}|^2}{a^2},\label{eq:tensorpowerspectrum}
\end{align}

where $\mathcal{P}^{(\epsilon)}_T(k)$ is the power spectrum for a single polarization of a tensor mode. The factor of $2$ accounts for both polarizations present in gravitational perturbations. For the modes that exit the Hubble horizon during inflation, the spectrum is evaluated at horizon exit, i.e. $k = aH$ \cite{MenaMarugan:2024tensor}. Since the amplitude of these modes is constant on superhorizon scales during inflation, it equals its value at the end of inflation, at which the spectra shown below are evaluated.

In the classical slow-roll limit, we obtain:
\begin{align}
    \mathcal{P}_T(k) = \frac{2 \kappa}{\pi^2} H^2,\label{eq:tensorpowerspectrum-classicalslowrolllimit}
\end{align}
where the Hubble rate is treated as approximately constant during slow-roll inflation. Thus, the classical treatment of cosmological perturbations predicts a scale-invariant regime in the tensor power spectrum at large $k$, and we expect the primordial tensor power spectra predicted by LQC, mLQC-I and mLQC-II to exhibit the same behavior for large values of $k$.

The scale-invariant amplitude of the tensor power spectrum alone is not a sufficient test of the predicted spectra, since this amplitude is not directly measured. Instead, other inflationary quantities can be computed for each model under the various treatments of cosmological perturbations, and these can be compared with observational data. We first use the same background solutions and initial states defined for all three models to compute the primordial scalar power spectra, defined as:
\begin{align}
    \mathcal{P}_S (k) = \frac{k^3}{2 \pi^2}\left|\frac{v_k}{z_S}\right|^2,\label{eq:scalarpowerspectrum}
\end{align}
where $v_k$ is the solution to the Mukhanov-Sasaki equation for scalar modes, and $z_S = a\dot{\phi}/H$ is the scalar pump field \cite{Li:2024lqc}. The resulting power spectra are then matched with the observable $A_s$, which according to the Planck Collaboration in the TT,TE,EE-lowE + lensing $68$\% limits data, has the value \cite{Planck:2020}:
\begin{align}
    \ln(10^{10} A_s) = 3.044 \pm 0.014 \hspace{2mm}(68\% \text{ CL}).
\end{align}

Using this, we fix the amplitude of the scalar power spectra to be $A_s$ i.e. $\mathcal{P}_{S}(k_\star) = A_s$ for some wavenumber $k_\star$ in units where $a_B=1$, which we identify with the pivot
scale $k_{\rm pivot}=k_\star/a_0=0.05\ {\rm Mpc}^{-1}$ (with $a_0$ the present scale
factor). The values of $k_\star$ for each model are given in
Table~\ref{TAB:kstar_table}.   In this way, each model with its own initial state has its own corresponding $k_\star$, which is then used to rescale all scalar and tensor power spectra. 
The wavenumbers are rescaled by $k_\star$, so that $k_\star$ corresponds to the pivot mode  and the scalar spectrum there equals $A_s$. The tensor power spectrum is rescaled with the same factors.
Once that is done, we compute the tensor-to-scalar ratio $r_\star$ defined as:
\begin{align}
    r_\star \equiv \frac{\mathcal{P}_T(k_\text{pivot})}{\mathcal{P}_{S}(k_\text{pivot})}.\label{eq:tensortoscalarratio}
\end{align}
The current observational upper bound on the tensor-to-scalar ratio is $r < 0.036$ at 95\% confidence \cite{BICEPKeckData}. The predicted $r_\star$ must satisfy this bound, which provides a consistency check on the normalized tensor spectra of each model in both the dressed metric and hybrid approaches.

\begin{table}[htbp]
  \centering
  \begin{tabular}{|w{c}{2cm}|w{c}{2.5cm}|w{c}{3cm}|w{c}{4cm}|}
    \hline
    \multicolumn{1}{|c}{} & & $k_\star$ $(N_e = 60)$ & $k_\star$ $(N_e = 65)$ \\
    \hline
    \multirow{2}{5em}{LQC} & \text{Hybrid} & $226.25-227.12$ & $33768.70-33795.71$ \\
    & \text{Dressed Metric} & $226.01 - 227.24$ & $33795.02-33806.52$ \\
    \hline
    \multirow{2}{5em}{mLQC-I} & \text{Hybrid} & $176.04$ & $26631.76$ \\
    & \text{Dressed Metric} & $175.77$ & $26658.73$ \\
    \hline
    \multirow{2}{5em}{mLQC-II} & \text{Hybrid} & $284.81-287.03$ & $42055.96-42060.19$ \\
    & \text{Dressed Metric} & $286.23-287.09$ & $42012.33-42051.76$ \\
    \hline
  \end{tabular}
  \captionsetup{justification=raggedright, singlelinecheck=false}
  \caption{\small The table shows $k_\star$ (in units where $a_B=1$) for all three LQC models, for both hybrid and dressed metric approaches. The ranges quoted for LQC and mLQC-II encompass the zeroth, second and fourth order adiabatic states. The single value quoted for mLQC-I corresponds to the initial state of each approach: the Hankel-function state for the hybrid approach and the de Sitter state for the dressed metric approach.}
  \label{TAB:kstar_table}
\end{table}

\begin{figure}
    \centering
    \includegraphics[width=1.0\linewidth]{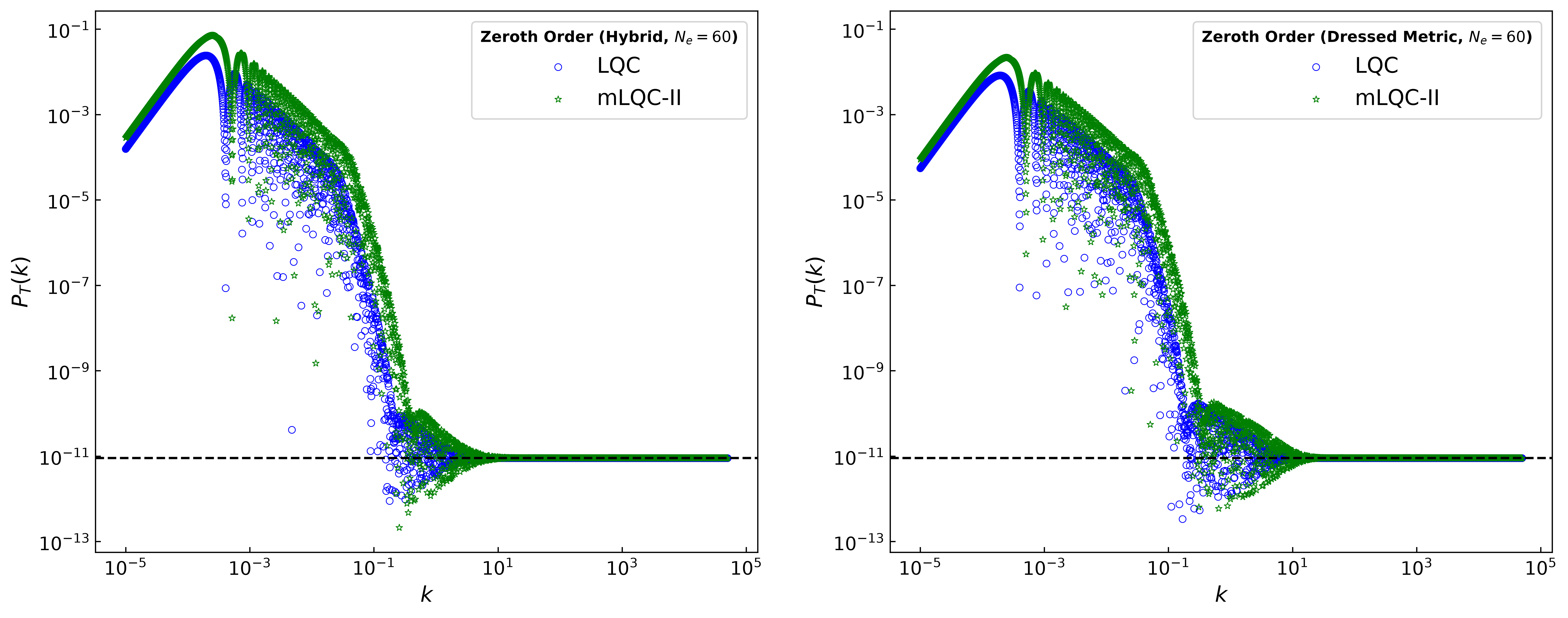}
    \includegraphics[width=1.0\linewidth]{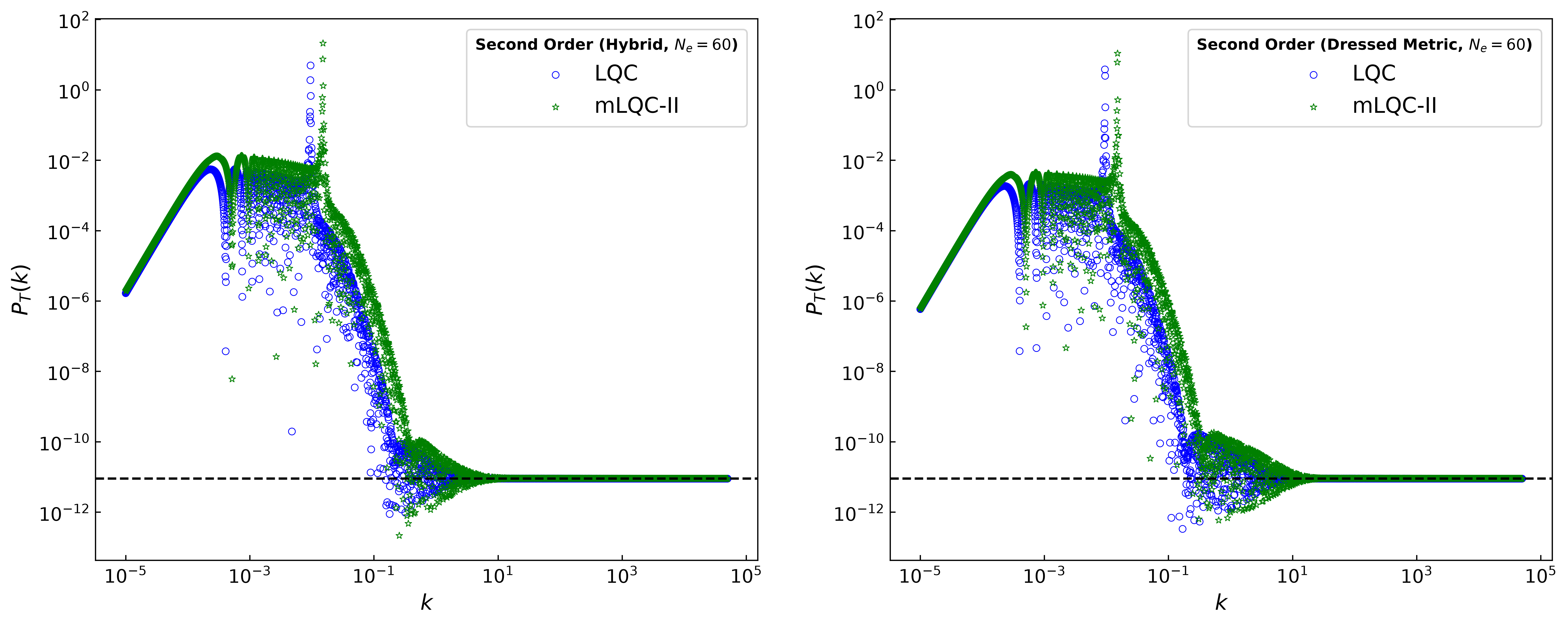}
    \includegraphics[width=1.0\linewidth]{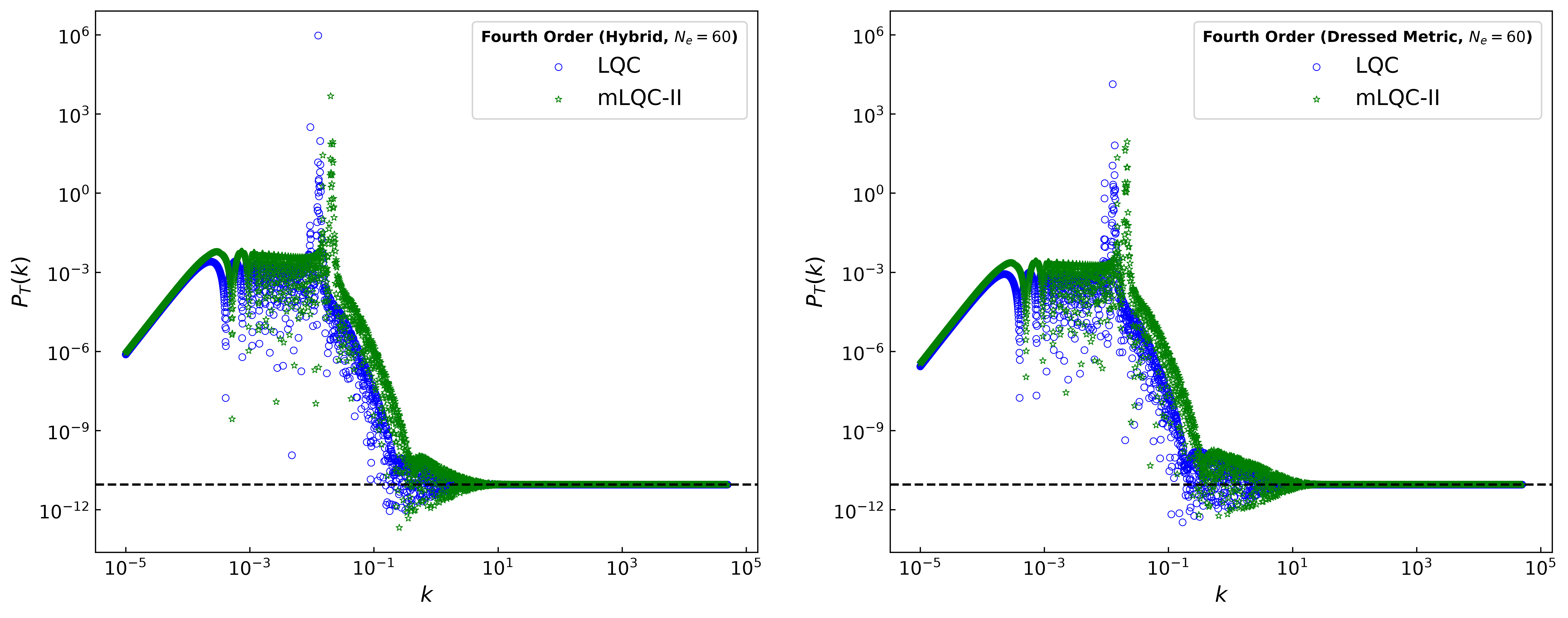}
    \captionsetup{justification=raggedright, singlelinecheck=false}
    \caption{\small The tensor power spectrum for LQC and mLQC-II with different adiabatic states (i.e. zeroth order in the first row, second order in the second row, and fourth order in the third row). The left column displays the power spectra in the hybrid approach, while the right side shows the dressed metric approach. $t_{\text{start}} = -1\times 10^{6}$ for all the initial states in both LQC and mLQC-II. In order to fix $N_e = 60$, we select $\phi_B = -1.433$ for LQC, and $\phi_B = -1.541$ for mLQC-II.}
    \label{FIG:LQC_mLQCII_ZOSOFO_HDM_60-efolds}
\end{figure}

\begin{figure}
    \includegraphics[width=1.0\linewidth]{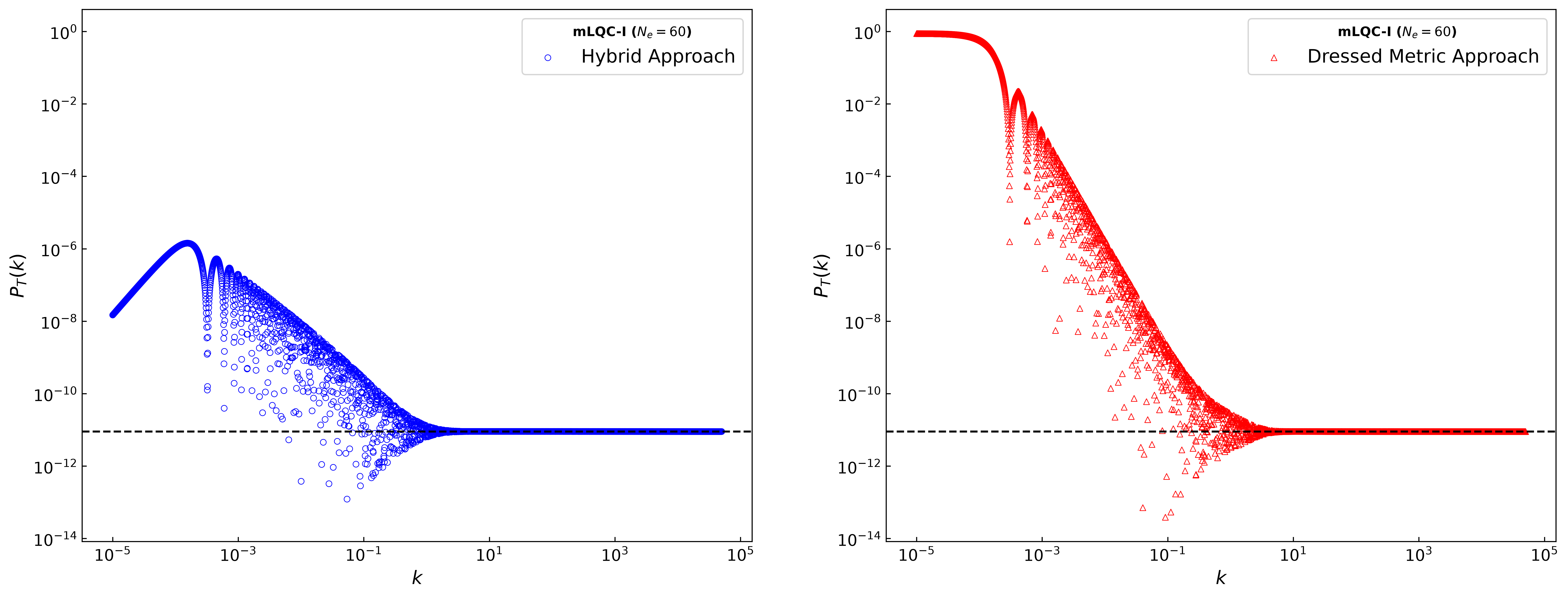}
    \captionsetup{justification=raggedright, singlelinecheck=false}
    \caption{\small Primordial tensor spectra for mLQC-I in hybrid (left) and dressed metric (right) approaches. The same $t_\text{start}$ value is imposed for both dressed metric and hybrid approaches i.e. $t_\text{start} = -12$. Here, $\phi_B = -1.314$ which fixes $N_e = 60$. The Hankel-function solution, Eq.~(\ref{eq:exactsol_tensormodes}), is used as the initial state in the hybrid approach. The de Sitter initial state is used for the dressed metric approach.}
    \label{FIG:tensorPS_mLQCI_60efolds}
\end{figure}

\subsection{Comparison of Tensor Power Spectra of all models}
With the given initial conditions for the background dynamics and the initial states for the linear perturbations, we numerically compute the tensor power spectrum in both the dressed metric and hybrid approaches for all three models, with $\phi_B$ that fixes $N_e = 60$, as given in Table \ref{TAB:phi_b_vals}. The dashed line in all plots shows the tensor amplitude at the pivot scale, $A_T = r_\star A_s$, obtained after the tensor power spectrum of each model is rescaled and $\mathcal{P}_{T}$ is evaluated at the pivot mode $k_{\text{pivot}} = 0.05$ Mpc$^{-1}$. It is approximately $A_T = 9.02 \times 10^{-12}$ for all models under both hybrid and dressed metric approaches.

In Fig. \ref{FIG:LQC_mLQCII_ZOSOFO_HDM_60-efolds}, tensor power spectra for LQC and mLQC-II under both hybrid and dressed metric approaches have been plotted for all three adiabatic initial states. In both approaches, one can observe that the amplification of the primordial power spectrum in the infrared regime depends on the initial states; more specifically, with the order of initial state getting higher, the amplitude of the tensor power spectrum is suppressed more. Furthermore, in the highly oscillatory regime, the amplitude has a downward trend for zeroth order but it remains somewhat steady for second and fourth order initial states. Notably, there is a spike in amplitude in the range $k \in (10^{-3},10^{-1})$ for both the second and fourth order initial states, which appears in both dressed metric and hybrid approaches. It must be emphasized that this is an effect of the choice of initial state: more specifically, $W_k$ vanishes for certain values of $k$ in each order, which causes the initial state to diverge, thereby generating a spike in our plots. Furthermore, for all three initial states, it can be seen that the tensor power spectra evaluated for mLQC-II all have larger amplitudes than LQC in both the infrared and amplified oscillatory regimes, but they approach the same scale invariant amplitude $A_T$ in the scale-invariant regime. This indicates that a difference in regularization approach once again can impact the primordial tensor power spectrum in significant ways. When comparing both approaches, it is observed that for the same order of adiabatic initial state, the amplitude of the tensor power spectrum is suppressed more in the dressed metric approach versus the hybrid approach in the infrared regime. However, in the amplified oscillatory regime, the amplitude becomes larger for the dressed metric approach. We therefore expect this difference to emerge in the angular power spectrum in the low multipole regime, which is most sensitive to amplitudes registered in the range of $k$ right next to the scale-invariant regime.

In Fig. \ref{FIG:tensorPS_mLQCI_60efolds}, we compare the mLQC-I tensor power spectra evaluated under the hybrid and dressed metric approaches. On the left, the hybrid approach uses the Hankel-function solution, Eq. (\ref{eq:exactsol_tensormodes}), as its initial state; on the right, the dressed metric approach uses Eq. (\ref{eq:deSittersolution}) as its initial state. Upon inspection, one immediately notices that the amplitudes in the infrared regime are much higher for the dressed metric approach versus the hybrid approach; additionally, the amplitude steadily rises in this regime for the hybrid approach, while it attains a plateau in the dressed metric approach, before both tensor power spectra steadily fall in the amplified oscillatory regime, and ultimately reach the same scale-invariant value $A_T$. Similar to the results shown in Fig. \ref{FIG:LQC_mLQCII_ZOSOFO_HDM_60-efolds}, here too we find that the amplitudes of the tensor power spectrum in the amplified oscillatory regime are higher for the dressed metric approach than the hybrid approach; once again, these differences will lead to significant effects in the angular power spectrum at low multipoles.

The infrared behavior obtained with the de Sitter initial state can be
understood analytically. In the emergent de Sitter phase of the
contracting branch, the scale factor behaves as $a = 1/(H_\Lambda \eta)$,
and the superhorizon solution of the mode equation with
$s^{(T)} \simeq -2/\eta^2$ satisfies $|h_k/a|^2 \simeq H_\Lambda^2/(2k^3)$. 
In the contracting de Sitter phase, the amplitude of each superhorizon mode is held fixed at a value determined entirely by the de Sitter Hubble rate.
Using
$H_\Lambda^2 = \kappa\alpha\rho_\Lambda/3$ with
$\rho_\Lambda = 3/[\kappa\alpha\lambda^2(1+\gamma^2)^2]$, the infrared
plateau of the tensor power spectrum is therefore predicted to be:
\begin{equation}
P_T(k) \simeq \frac{2\kappa H_\Lambda^2}{\pi^2}
= \frac{2\kappa}{\pi^2\lambda^2(1+\gamma^2)^2},
\label{eq:plateau}
\end{equation}
which is independent of the inflaton mass $m$ and of $\phi_B$. The
plateau amplitude is set by the Planckian de Sitter rate and lies about
eleven orders of magnitude above the inflationary value $A_T$,
consistent with the spectra shown in Fig. \ref{FIG:tensorPS_mLQCI_60efolds}. In contrast, the Hankel-function state used in the hybrid approach is sensitive to $t_{\text{start}}$ and does not reproduce this constant-amplitude superhorizon solution of the contracting
de Sitter branch, which is why the hybrid infrared spectrum is suppressed relative to the dressed metric one.

We are interested in seeing how increasing the number of inflationary e-foldings from $N_e = 60$ to $N_e = 65$ changes the tensor power spectra. To test this, we compute the tensor power spectra with the same initial conditions as before, except that we use $\phi_B$ values corresponding to $N_e = 65$. Then, we use the same rescaling approach that has been outlined in prior sections to compute the values of $k_\star$. For the sake of comparison, data for $k_\star$ has been presented for both $N_e = 60$ and $N_e = 65$ in Table \ref{TAB:kstar_table}. It can be seen that across both numbers of inflationary e-foldings, $k_\star$ is largest for mLQC-II and smallest for mLQC-I, for both hybrid and dressed metric approaches. Furthermore, $k_\star$ increases drastically as the number of inflationary e-foldings increases. This is expected, because a larger $N_e$ means more expansion between the bounce and the horizon exit of the pivot mode, so that $a(t_\star)$, and hence $k_\star = a(t_\star)H(t_\star)$, is larger. It must be emphasized that despite the increase in the number of e-foldings, the qualitative results in the infrared and amplified oscillatory regimes do not differ from their $N_e = 60$ counterparts. Increasing $N_e$ therefore acts as a robustness check, in which the trends displayed by the tensor power spectra do not change drastically.

\begin{table}[htbp]
  \centering
  \begin{tabular}{|w{c}{2cm}|w{c}{2.5cm}|w{c}{4.5cm}|w{c}{4.5cm}|}
    \hline
    \multicolumn{1}{|c}{} & & $r_\star$ $(N_e = 60)$ & $r_\star$ $(N_e = 65)$ \\
    \hline
    \multirow{2}{5em}{LQC} & \text{Hybrid} & $0.00429729-0.00429743$ &  $0.00429725-0.00429741$\\
    & \text{Dressed Metric} & $0.00429689-0.00429813$ & $0.00429735-0.00429753$ \\
    \hline
    \multirow{2}{5em}{mLQC-I} & \text{Hybrid} & $0.00429752$ & $0.00429735$ \\
    & \text{Dressed Metric} & $0.00429705$ & $0.00429720$ \\
    \hline
    \multirow{2}{5em}{mLQC-II} & \text{Hybrid} & $0.00429708-0.00429780$ & $0.00429721-0.00429753$ \\
    & \text{Dressed Metric} & $0.00429735-0.00429765$ & $0.00429746-0.00429751$ \\
    \hline
  \end{tabular}
  \captionsetup{justification=raggedright, singlelinecheck=false}
  \caption{\small $r_\star$ for all three LQC models, for both hybrid and dressed metric approaches. The ranges quoted for LQC and mLQC-II encompass the zeroth, second and fourth order adiabatic states. The single value quoted for mLQC-I corresponds to the initial state of each approach: the Hankel-function state for the hybrid approach and the de Sitter state for the dressed metric approach.}
  \label{TAB:r_star_list}
\end{table}

\subsection{Tensor-to-Scalar Ratio}

As discussed before, the tensor-to-scalar ratio $r$ is an important inflationary observable that must be computed and checked before computing the angular power spectra for each model under various treatments of perturbations. This ratio is computed using Eq. (\ref{eq:tensortoscalarratio}), once the values of the comoving wavenumber $k_\star$ are computed for each model for all initial states. This is done for both $N_e = 60$ and $N_e = 65$, which allows us to test the sensitivity of our tensor power spectra and corresponding inflationary observables to the number of inflationary e-foldings.

From the data presented, it can be seen that all three models under both the hybrid and dressed metric approaches predict a tensor-to-scalar ratio of $r_\star = 4.3 \times 10^{-3}$, for both $N_e = 60$ and $N_e = 65$, at the time when the wavemode crosses the Hubble horizon. This value satisfies the current bound $r<0.036$ \cite{BICEPKeckData}. The fact that each model predicts the same $r_\star$ suggests that, for these two values of $N_e$, this predicted value does not depend on the regularization, specifically for an inflationary universe that is driven by the Starobinsky potential. It is also insensitive to the approach used to treat cosmological perturbations, and it changes very little between $N_e = 60$ and $N_e = 65$.
It is to be noted that the equality of $r_\star$ for $N_e=60$ and $65$ follows from the normalization. In classical 
slow-roll inflation $r$ is set by the number of e-folds $N_\star$ between the exit of the
pivot mode from the Hubble radius and the end of inflation. For a fixed inflaton mass, the condition
$\mathcal{P}_S(k_\star)=A_s$ fixes $N_\star$. Solving the slow-roll equations of general
relativity for the same $m$ and $A_s$ gives $r=4.30\times10^{-3}$, which agrees with
Table~\ref{TAB:r_star_list}. Increasing $N_e$ by five adds five e-folds before the pivot
mode exits, so $k_\star$ increases by $e^5$ (Table~\ref{TAB:kstar_table}) while $N_\star$
and $r_\star$ do not change. The value of $r_\star$ is therefore the standard inflationary
one since the pivot mode exits long after the bounce, in the scale-invariant regime. A
different inflaton mass would change $N_\star$ and with it $r_\star$. In this manuscript, we have kept $m$ fixed.

%It must be noted that we have tested only these two values, so we do not claim that $r_\star$ is independent of $N_e$ in general. This is expected because $r_\star$ is evaluated in the scale invariant regime, wherein all models agree on the same values of $A_s$ and $A_T$. Table \ref{TAB:r_star_list} is thus further confirmation that all three models tend to the same classical results at large $k$.

\subsection{Scale Dependence of the Tensor-to-Scalar Ratio}

\begin{figure}
    \centering
    \includegraphics[width=0.8\linewidth]{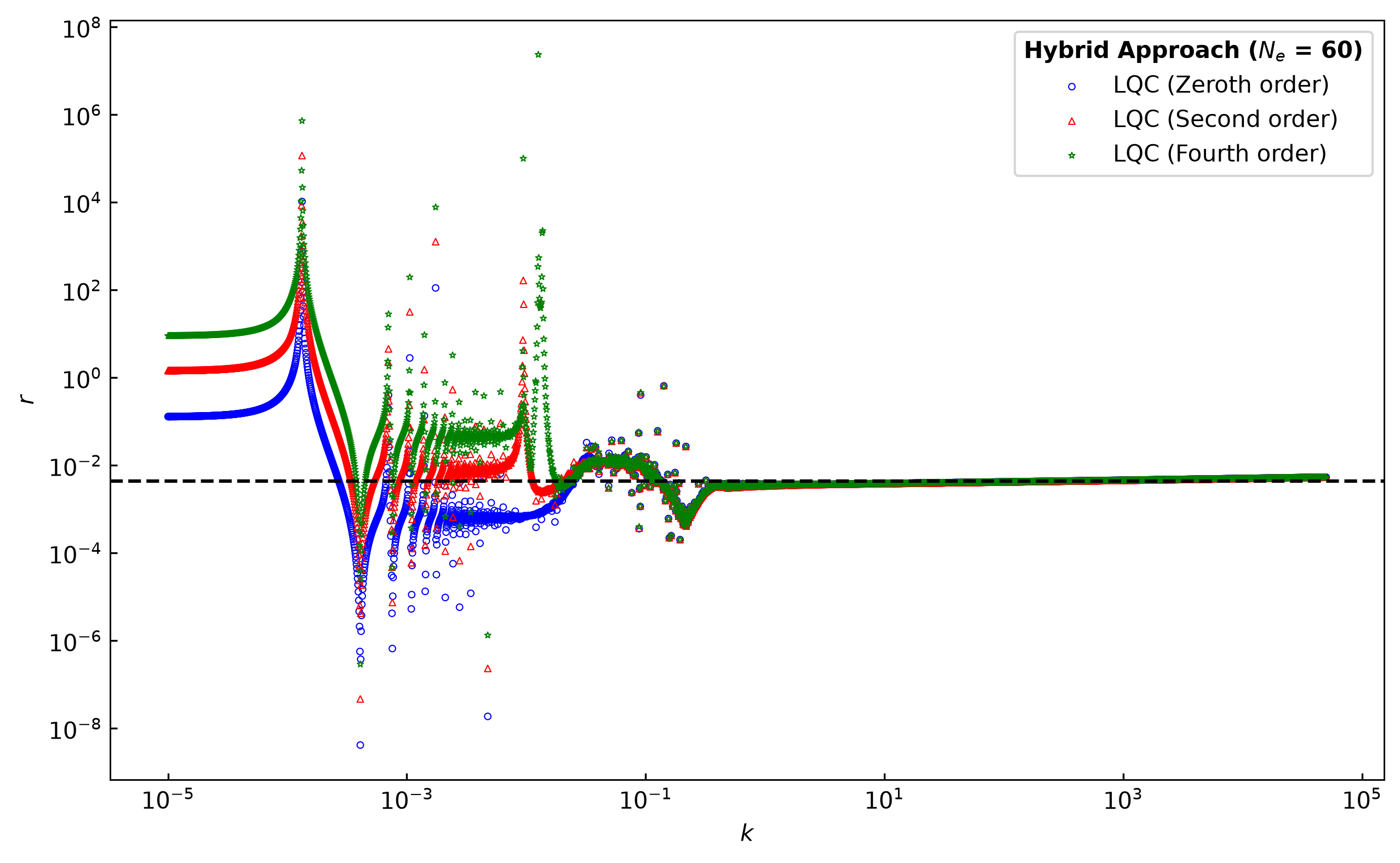}
    \includegraphics[width=0.8\linewidth]{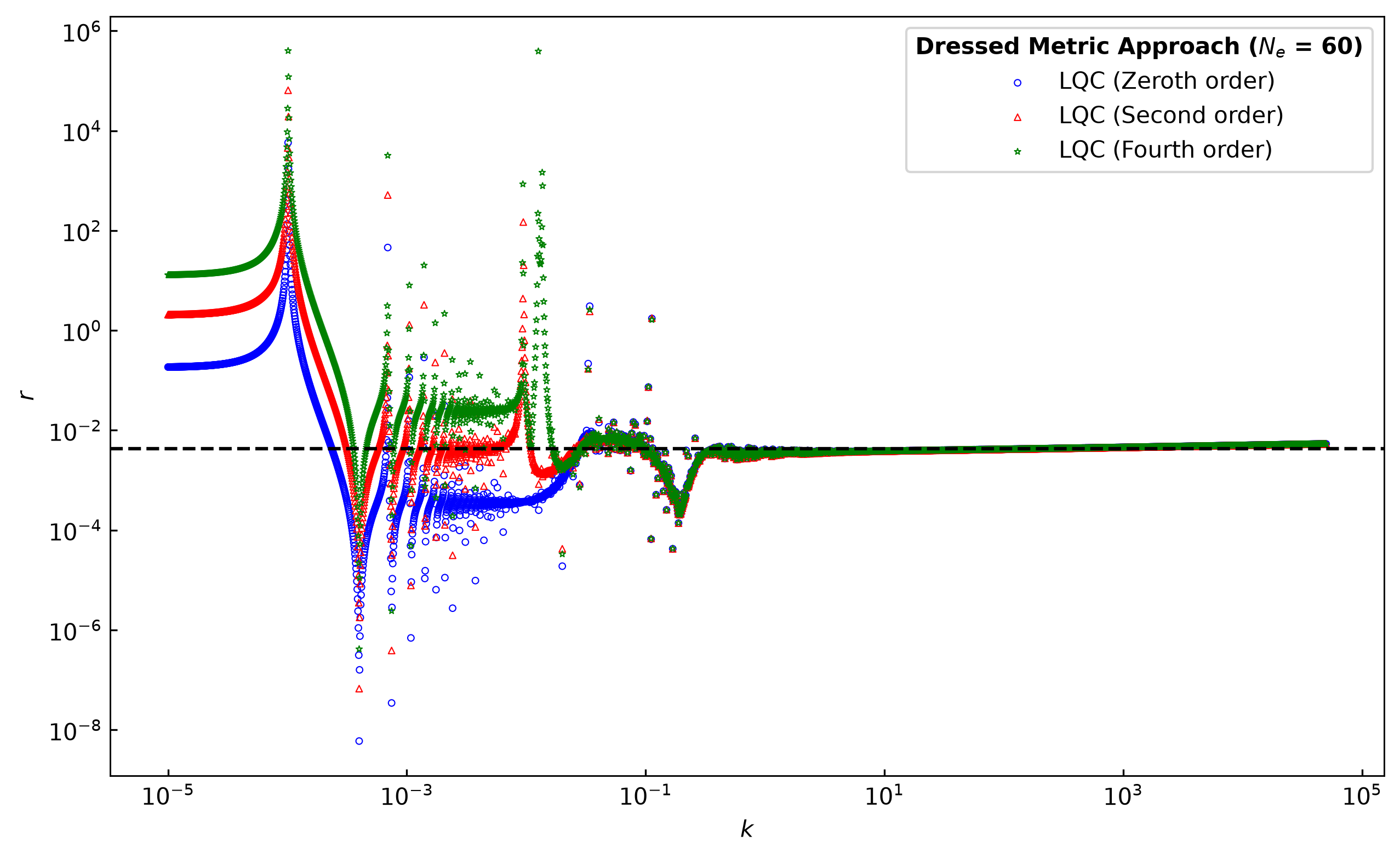}
    \captionsetup{justification=raggedright, singlelinecheck=false}
    \caption{\small Plot of $r$ vs. $k$ for standard LQC using the adiabatic initial states. The graph at the top displays results under the hybrid approach, while the graph at the bottom displays results under the dressed metric approach. The number of inflationary e-foldings chosen is $N_e = 60$. The dashed line is $r_\star$. Isolated spikes for the second and fourth order states come from the zeros of $W_k$ (Sec.~V).}
    \label{FIG:r_vs_k_LQC_HDM_60efolds}
\end{figure}

\begin{figure}
    \centering
    \includegraphics[width=0.8\linewidth]{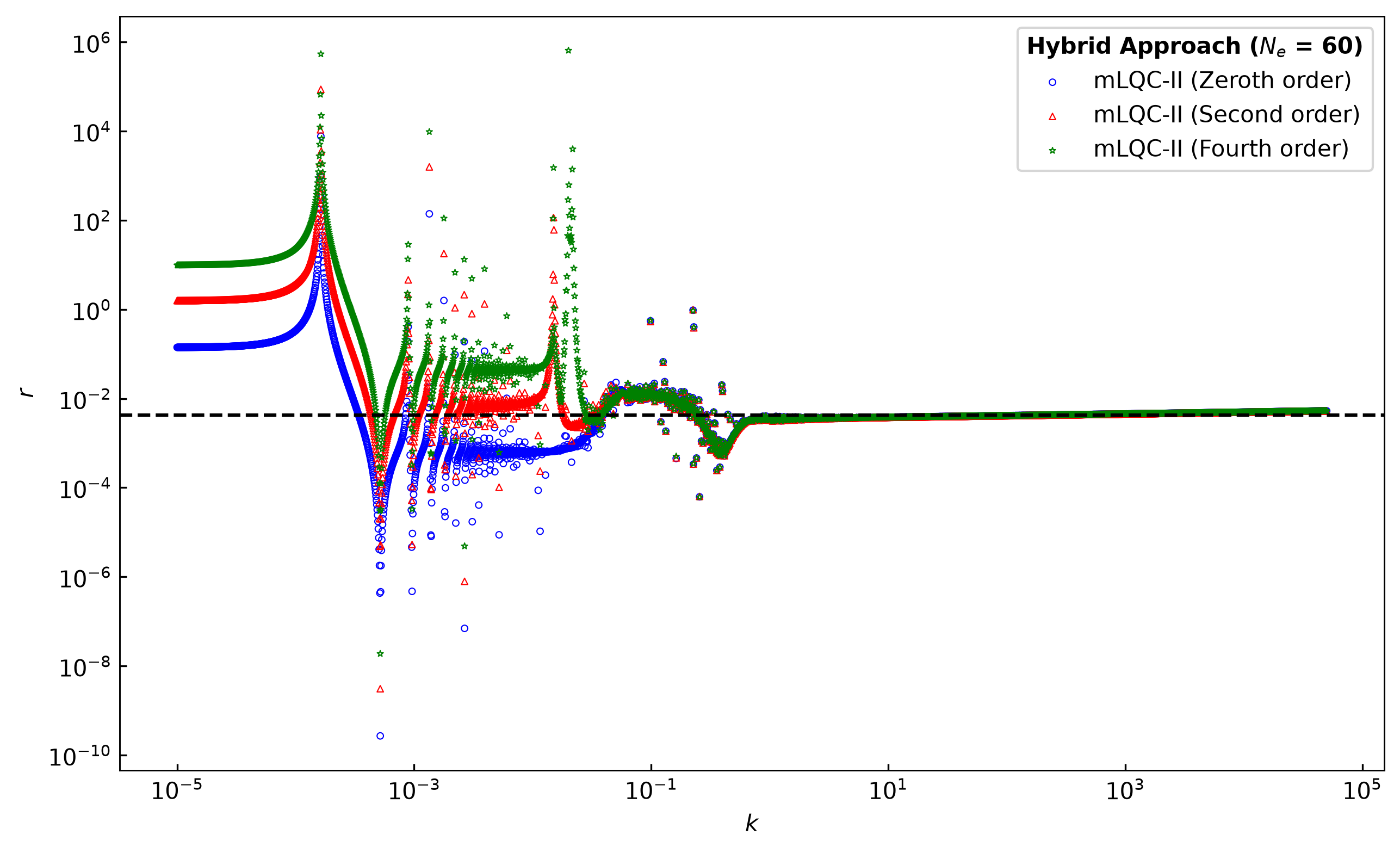}
    \includegraphics[width=0.8\linewidth]{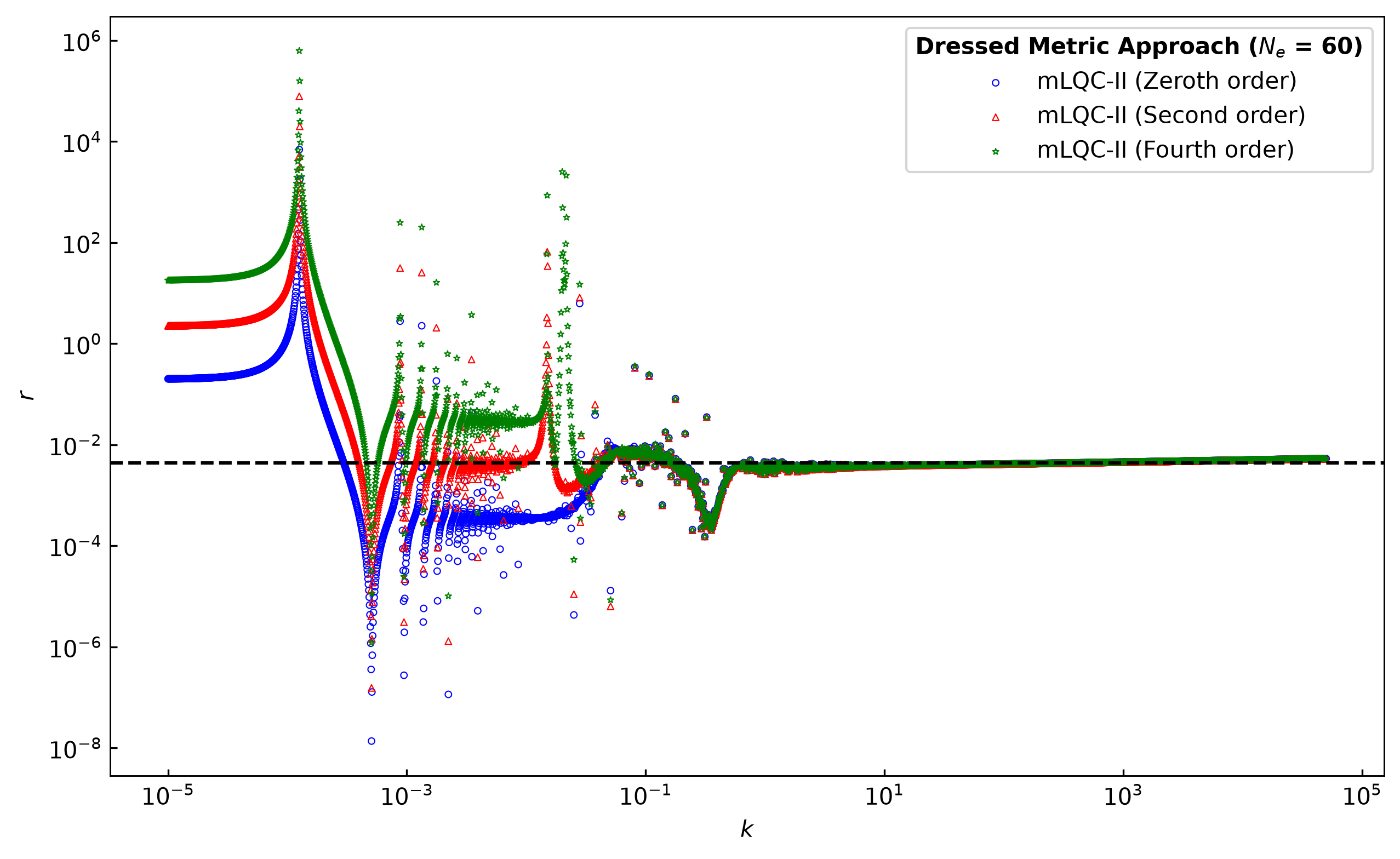}
    \captionsetup{justification=raggedright, singlelinecheck=false}
    \caption{\small Plot of $r$ vs. $k$ for mLQC-II using the adiabatic initial states. The graph at the top displays results under the hybrid approach, while the graph at the bottom displays results under the dressed metric approach. The number of inflationary e-foldings chosen is $N_e = 60$. The dashed line is $r_\star$. Isolated spikes for the second and fourth order states come from the zeros of $W_k$ (Sec.~V).}
    \label{FIG:r_vs_k_mLQCII_HDM_60efolds}
\end{figure}

\begin{figure}
    \centering
    \includegraphics[width=0.8\linewidth]{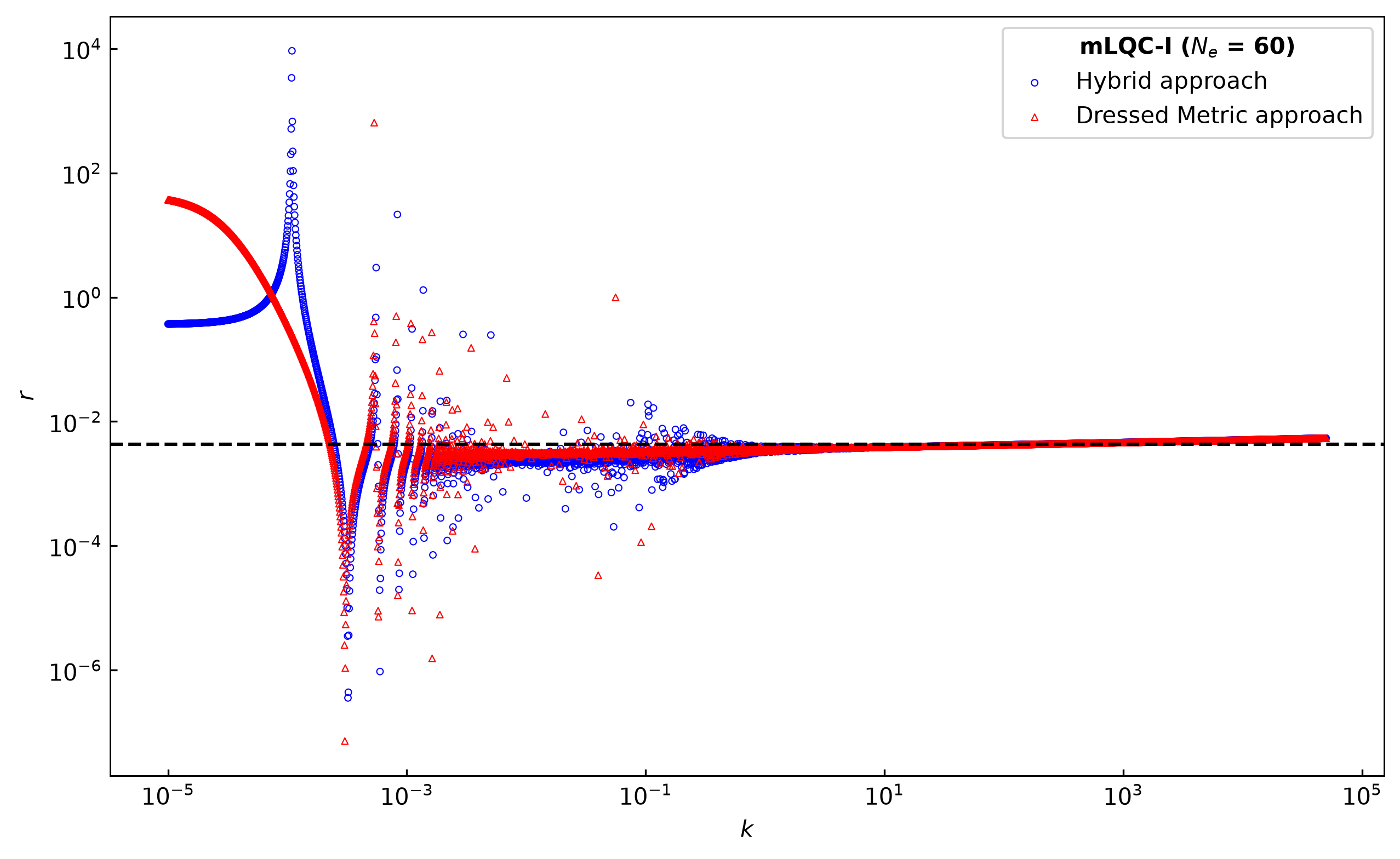}
    \captionsetup{justification=raggedright, singlelinecheck=false}
    \caption{\small Plot of $r$ vs. $k$ for mLQC-I, where $N_e = 60$. For the hybrid approach, the Hankel-function solution, Eq.~(\ref{eq:exactsol_tensormodes}), is used as the initial state, while for the dressed metric approach, the de Sitter initial state is used. The dashed line is $r_\star$.}
    \label{FIG:r_vs_k_mLQCI_HDM_60efolds}
\end{figure}

So far, the tensor-to-scalar ratio has been quoted only at the pivot scale. It must be noted that $r$ is in fact a function of $k$,
\begin{equation}
    r(k) = \frac{\mathcal{P}_T(k)}{\mathcal{P}_S(k)},
\end{equation}
and that $r_\star = r(k_\star)$ is its value at a single point. In Figs. \ref{FIG:r_vs_k_LQC_HDM_60efolds}, \ref{FIG:r_vs_k_mLQCII_HDM_60efolds} and \ref{FIG:r_vs_k_mLQCI_HDM_60efolds} we plot $r(k)$ for LQC, mLQC-II and mLQC-I respectively, under both the hybrid and dressed metric approaches and with $N_e = 60$. The dashed line marks $r_\star$. Since both spectra are rescaled with the same normalization, $r_\star = A_T/A_s$, and we can write:
\begin{equation}
    \frac{r(k)}{r_\star} = \frac{\mathcal{P}_T(k)/A_T}{\mathcal{P}_S(k)/A_s}.
\end{equation}
Thus $r(k)/r_\star$ compares how much the tensor and the scalar modes are enhanced or suppressed with respect to their scale-invariant amplitudes, and any effect that is the same for both sectors cancels in this ratio. What remains is the difference in how the two sectors respond to the bounce and to the pre-inflationary evolution.

In the scale-invariant regime, $r(k)$ is nearly constant, but it is not exactly constant. It grows slowly with $k$, by roughly twenty percent between $k_\star$ and the largest wavenumbers shown. This is the expected slow-roll behavior. For modes that leave the Hubble radius during inflation, $\mathcal{P}_T = 2\kappa H^2/\pi^2$ and $\mathcal{P}_S = \kappa H^2/(8\pi^2\varepsilon)$, so that $r = 16\,\varepsilon$, which gives to the leading order $r \simeq 12/N^2$ for the Starobinsky potential, where $N$ is the number of e-folds between horizon exit and the end of inflation. Modes with larger $k$ exit the horizon later, so $N$ is smaller and $r$ is larger. Equivalently, $d\ln r/d\ln k = n_T - (n_s - 1)$, where $n_T$ and $n_s$ are the tensor and scalar spectral indices. This is positive since $n_s < 1$ and $|n_T|$ is much smaller than $1-n_s$. In this regime $r(k)$ is therefore determined by the inflationary phase alone, and it is the
same for all models, both approaches and all initial states. This is why $r_\star$ in
Table~\ref{TAB:r_star_list} is the same for all of them. The pivot mode exits the Hubble
radius the same number of e-folds $N_\star$ before the end of inflation in every case, and
$r_\star = r(k_\star)$ (Sec.~V C).

In the upper part of the amplified oscillatory regime, here for $k \sim 0.05$ to $0.3$, $r(k)$ oscillates around $r_\star$. It rises above $r_\star$ by a factor of a few and then falls below it by a similar factor. For LQC and mLQC-II, the zeroth, second and fourth order curves lie on top of each other here in both approaches (Figs. \ref{FIG:r_vs_k_LQC_HDM_60efolds} and \ref{FIG:r_vs_k_mLQCII_HDM_60efolds}). The reason is that these modes are well inside the curvature scale at the initial time, so the adiabatic condition is satisfied and the result does not depend on the order of the adiabatic state. This makes the oscillation a genuine imprint of the bounce and not of the initial state. More specifically, the scalar and the tensor modes obey Eq. (\ref{eq:MSequation-tensormodes}) with different effective masses, since the scalar mass contains terms that depend on the inflaton potential and on the coupling of the inflaton perturbation to the metric, which the tensor mass does not. The two sectors therefore acquire different oscillation phases while passing through the bounce, the oscillations in $\mathcal{P}_T$ and $\mathcal{P}_S$ do not coincide, and $r(k)$ oscillates around $r_\star$. If the two sectors were modified in the same way, $r(k)$ would be equal to $r_\star$ at all $k$.

For $k \sim 10^{-3}$ to $10^{-2}$, the points scatter over several orders of magnitude. This is due to the fast oscillation of the two spectra, which is already seen in $\mathcal{P}_T$ in Fig. \ref{FIG:LQC_mLQCII_ZOSOFO_HDM_60-efolds}, so only the envelope of the scatter carries information. The isolated spikes of the second and fourth order states, which reach values far above $r_\star$, come from the zeros of $W_k$ discussed before. The same spikes appear in $\mathcal{P}_T$ in Fig. \ref{FIG:LQC_mLQCII_ZOSOFO_HDM_60-efolds} and they grow with the adiabatic order. They are a property of the initial state, and should not be interpreted as a prediction of the model.

At smaller $k$ the curves become smooth again. There is a peak near $k \approx 1.3 \times 10^{-4}$, followed by a deep trough near $k \approx 4 \times 10^{-4}$. These positions are the same for all adiabatic orders, for both approaches, and for LQC and mLQC-II, while only the height of the peak and the depth of the trough change with the order. The positions are therefore set by the background, and the amplitudes by the initial state. The trough coincides with the first minimum of $\mathcal{P}_T$ in Fig. \ref{FIG:LQC_mLQCII_ZOSOFO_HDM_60-efolds}, so it comes from the tensor sector. At the peak, $\mathcal{P}_T$ is smooth, so the peak presumably comes from a minimum of $\mathcal{P}_S$. In this way, $r(k)$ shows where the first nodes of the scalar and tensor oscillations fail to coincide, which is not easy to see in either spectrum alone. In the infrared regime, $r$ is larger than $r_\star$ by one to several orders of magnitude in every case, so the tensor modes are enhanced more than the scalar modes there. However, the value of $r$ at the smallest $k$ changes by orders of magnitude with the adiabatic order (for example, from about $0.1$ to about $10$ in the hybrid approach for LQC). It is therefore a feature of the chosen state and not a prediction of the model.

When comparing the models and the approaches, the curves for LQC and mLQC-II have the same structure (Figs. \ref{FIG:r_vs_k_LQC_HDM_60efolds} and \ref{FIG:r_vs_k_mLQCII_HDM_60efolds}). The peak and the trough are at nearly the same wavenumbers, and the main difference is in the amplitudes. For both models, the hybrid and the dressed metric panels look almost the same, although $\mathcal{P}_T$ itself differs between the approaches (Fig. \ref{FIG:LQC_mLQCII_ZOSOFO_HDM_60-efolds}). This implies that going from one approach to the other changes $\mathcal{P}_S$ in much the same way as $\mathcal{P}_T$, and the ratio removes most of this dependence. The case of mLQC-I is different, as can be seen from Fig. \ref{FIG:r_vs_k_mLQCI_HDM_60efolds}; here, each approach has a single initial state, and the two curves differ at small $k$. In the hybrid approach $r$ starts at a value of order $0.3$ and rises to a peak near $k \approx 10^{-4}$, whereas in the dressed metric approach it starts at a value of order tens, is flat at the smallest $k$, and decreases without a peak. Both curves have a trough near $k \approx 3 \times 10^{-4}$, and both settle on $r_\star$ by $k \sim 1$ with small oscillations. For mLQC-I, the infrared difference between the two approaches is therefore not removed by the ratio, which agrees with the different effective masses and initial states of the two approaches in the contracting branch.

To summarize, $r(k)$ is independent of the initial state only in the scale-invariant regime, where it follows the slow-roll behavior, and in the upper part of the amplified oscillatory regime, where it oscillates around $r_\star$ in a way that does not depend on the order of the adiabatic state. The peaks, troughs and spikes at smaller $k$ depend on the initial state. It must also be noted that the infrared part of $r(k)$ cannot be observed directly. In the CMB it can only be seen through the lowest multipoles, where cosmic variance and foregrounds are large. As discussed in the next section, the low multipoles are most sensitive to the modes in the upper part of the amplified oscillatory regime and next to the scale-invariant regime. These are the modes where $r(k)$ differs from $r_\star$ by a factor of a few and where the dependence on the initial state starts. This is why the differences among the models in the $B$-mode spectrum are confined to $\ell \lesssim 20$, and why a dependence on the initial state also appears there, in the difference between the second and fourth order hybrid results.

\section{CMB $B$-Mode Predictions}

\begin{figure}
    \centering
    \includegraphics[width=0.8\linewidth]{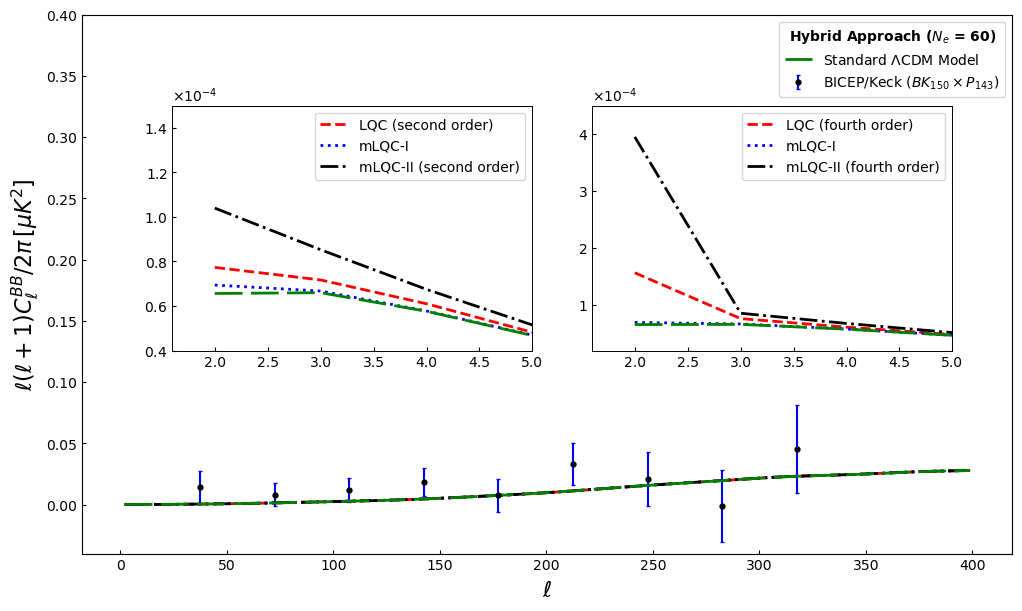}
    \captionsetup{justification=raggedright, singlelinecheck=false}
    \caption{\small The angular power spectrum for $C^{BB}_\ell$ predicted by all three quantum cosmological models under the hybrid approach with $N_e = 60$. The inset plot on the left uses second order adiabatic initial states for LQC and mLQC-II, while the Hankel-function solution, Eq.~(\ref{eq:exactsol_tensormodes}), is used as the initial state for mLQC-I. On the other hand, the inset plot on the right uses fourth order adiabatic initial states for LQC and mLQC-II, while the same Hankel-function state is used for mLQC-I. The black dots represent the BICEP/Keck 2018 BB angular power spectra for the bandpower $BK_{150} \times P_{143}$, with blue error bars for all multipoles. The green curve is the lensed $\Lambda$CDM+$r$ best fit to the BICEP/Keck 2018 data, computed for $r=4.3 \times 10^{-3}$ and includes lensed $B$-modes.}
    \label{FIG:Cbb_Hybrid_60_FO+DS+FO}
\end{figure}

\begin{figure}
    \centering
    \includegraphics[width=0.8\linewidth]{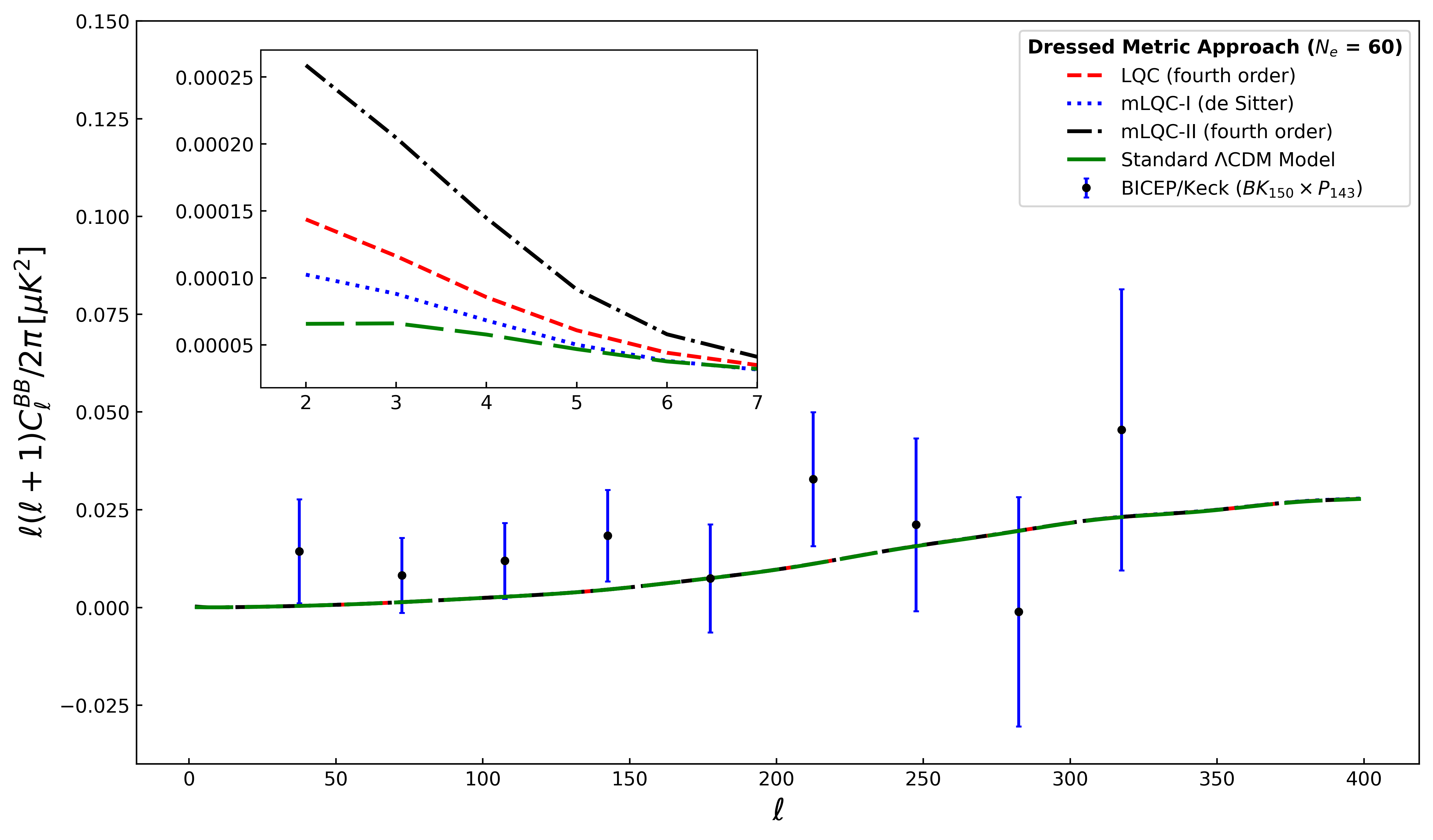}
    \captionsetup{justification=raggedright, singlelinecheck=false}
    \caption{\small The angular power spectrum for $C^{BB}_\ell$ predicted by all three quantum cosmological models under the dressed metric approach with $N_e = 60$. Fourth order adiabatic initial states are used for LQC and mLQC-II, while the de Sitter initial state is used for mLQC-I. The black dots represent the BICEP/Keck 2018 BB angular power spectra for the bandpower $BK_{150} \times P_{143}$, with blue error bars for all multipoles. The green curve is the lensed $\Lambda$CDM+$r$ best fit to the BICEP/Keck 2018 data, computed for $r=4.3 \times 10^{-3}$ and includes lensed $B$-modes.}
    \label{FIG:Cbb_DressedMetric_60_FO+DS+FO}
\end{figure}

After presenting the tensor power spectra for all three LQC models and discussing the differences between the hybrid and dressed metric approaches for various choices of initial states, we now compute the corresponding $B$-mode angular power spectra, which are the cleanest probe of the tensor perturbations. All the plots are compared with the lensed $\Lambda$CDM+$r$ prediction for $r=4.3\times10^{-3}$ and with the BICEP/Keck 2018 $C^{BB}_\ell$ bandpowers \cite{BICEPKeckData}. We first outline how the tensor power spectra computed in the previous section are fed into \textsc{CAMB}. We then present the results, in which adiabatic initial states are used for LQC and mLQC-II, while the initial state used for mLQC-I depends on the perturbative approach: the Hankel-function solution, Eq.~(\ref{eq:exactsol_tensormodes}), in the hybrid approach and the de Sitter initial state, Eq.~(\ref{eq:deSittersolution}), in the dressed metric approach. Our primary focus is on deviations at large angles (i.e. the small $\ell$ regime). Within each perturbative approach, such deviations reflect differences between the regularizations, and could in principle indicate which regularization is favored by the data. We also compare the two approaches to the treatment of cosmological perturbations.

\subsection{From Primordial Tensor Power Spectrum to $C_\ell^{BB}$}

When computing the angular power spectrum, our primary focus is on the $B$-mode angular power spectrum, represented by $C^{BB}_\ell$. This is because the detection of a $B$-mode polarization signal points to the existence of primordial gravitational waves. Any effects arising from tensor perturbations in the early universe must therefore be reflected in $C^{BB}_\ell$ as a function of the multipole $\ell$, especially considering that the effects of the pre-inflationary phase are encoded in the initial states and the effective mass functions that enter Eq. (\ref{eq:MSequation-tensormodes}). Before the spectra are passed to \textsc{CAMB}, an average over 20 sample points is taken to prevent the fast oscillations in the tensor power spectrum from affecting the \textsc{CAMB} results. Then all tensor power spectra are rescaled with the same
normalization factor fixed by the scalar matching, so that the rescaled tensor
power spectrum at the pivot scale is $\mathcal{P}_{T}(k_\star) = r_\star A_s = A_T.$
%Then rescaling is done of all tensor power spectra such that $\mathcal{P}_{T}(0.05)=A_s$ as described before. 
This allows the model predictions to be fairly compared with observational data. Finally, we feed the rescaled tensor power spectra into \textsc{CAMB}, wherein only the $B$-mode results are taken into account.

We use the standard transfer function because the background dynamics of all three models rapidly approach the general relativistic trajectory as the curvature falls below the Planckian scale, and by the time the observable modes exit the horizon during inflation the effective spacetime is indistinguishable from the classical one. The entire model dependence of the calculation therefore enters through the primordial tensor power spectrum computed in Sec.~V, which carries the information about the pre-inflationary evolution, the choice of initial state, and the effective tensor mass in the neighborhood of the bounce. This treatment parallels the one adopted for the scalar sector \cite{Li:2024lqc}, so that the scalar and tensor predictions of the two analyses can be compared within a common framework. It also implies that the conclusions drawn here are robust with respect to the details of the late-time cosmology: a change in the cosmological parameters or in the transfer function affects all models in the same way and does not modify the relative comparison among the regularizations.

The transfer function computed by \textsc{CAMB}~\cite{Lewis:1999bs} is used to model standard cosmology, with the fixed parameters: the Hubble parameter at present $H_0=67.36$ km s$^{-1}$ Mpc$^{-1}$, physical baryon density parameter $\Omega_bh^2 = 0.02237$, physical cold dark matter density parameter $\Omega_ch^2=0.1200$, the sum of neutrino masses in electronvolts $m_\nu=0.06$, curvature parameter $\Omega_k=0$ (indicating a flat universe) and optical depth to reionization $\tau=0.0544$.

The black dots represent the $B$-mode signals detected by BICEP/Keck during the 2018 observing season \cite{BICEPKeckData}. Out of the various bandpowers that were available, we selected $\text{BK}_{150} \times \text{P}_{143}$ for three main reasons: firstly, a cross-correlation between two different maps, i.e. BICEP/Keck ($BK$) and Planck ($P$), suppresses instrumental noise; secondly, both channels are near the CMB frequency peak; and finally, dust and synchrotron contamination are minimal. It is expected that modifications to the amplified oscillatory regime of the primordial power spectrum should lead to visible effects in the angular power spectrum at large angles, i.e. low $\ell$ multipoles \cite{Li:2024lqc}.

\subsection{Model Comparison}

The results for the angular $BB$ power spectra are shown in Figs. \ref{FIG:Cbb_Hybrid_60_FO+DS+FO}-\ref{FIG:Cbb_DressedMetric_60_FO+DS+FO} for both hybrid and dressed metric approaches, and for various initial quantum states. In all the plots, we use the adiabatic initial states for LQC and mLQC-II, while we use different initial states for each perturbative approach when modeling results for mLQC-I. It can be seen from all the plots that all three models match the best fit curve at large multipoles $\ell$ in both the hybrid and dressed metric approaches; this is expected since their tensor power spectra predict the same scale-invariant trend for large wavenumbers. However, all three models show visible deviations at low multipoles, which are worth exploring.

In Fig. \ref{FIG:Cbb_Hybrid_60_FO+DS+FO}, the $B$-mode spectrum for all three models is presented under the hybrid approach, with $N_e = 60$. We show the second and fourth order initial states for LQC and mLQC-II, and the Hankel-function solution for mLQC-I, because the results for the zeroth order initial state are identical to those for the second order initial state. One can see that at low values of $\ell$, mLQC-II shows the largest deviation from the standard model, while mLQC-I deviates the least across both orders of adiabatic initial states. This can be attributed to the fact that in the upper part of the amplified oscillatory regime and at the start of the scale-invariant regime, where $k \in (0.01,10)$, the amplitudes of mLQC-II are always larger than both LQC and mLQC-I. Furthermore, when comparing the fourth order adiabatic results with the second order ones, it can be seen that despite the same qualitative trend between all three models, the deviations are larger for fourth order adiabatic initial states than for second order ones. This can be attributed to the fact that as the order of the initial state increases, the amplitudes of the tensor power spectra in the infrared regime are strongly suppressed, while the amplitudes increase slightly in the range of $k$ right next to the scale-invariant regime. The dependence on the adiabatic order seen here is a dependence on the initial state, and is not a prediction of the model.

The angular $BB$ power spectra have also been plotted for the three models under the dressed metric approach in Fig. \ref{FIG:Cbb_DressedMetric_60_FO+DS+FO}. This time, the results are displayed for the fourth order adiabatic initial state for LQC and mLQC-II, because they are identical to the results for the zeroth and second order initial states. For mLQC-I, we use the de Sitter initial state. Here too, we find the same qualitative result as before: mLQC-II shows the largest divergence away from the standard $\Lambda$CDM model, while mLQC-I shows the least deviation. This too can be linked with the amplitudes of the tensor power spectra for each model in the amplified oscillatory regime; even under the dressed metric approach, mLQC-II has the larger amplitudes, while mLQC-I has the smallest. Upon comparing the angular power spectra between the hybrid and dressed metric approaches, one can clearly see that the deviations displayed by the dressed metric approach are much larger than those visible for the hybrid approach. This can be attributed to the trend observed when comparing primordial tensor power spectra across the two approaches for each model: from the infrared up to the amplified oscillatory regime, the primordial tensor power spectra computed under the dressed metric approach always have larger amplitudes than those of the hybrid approach.

\subsection{Observational Interpretation}

The $B$-mode spectra computed above can be placed in the context of the planned LiteBIRD mission, a full-sky CMB polarization satellite whose primary science goal is the detection of the imprint of inflationary gravitational waves, and which is designed to measure the tensor-to-scalar ratio with a total uncertainty $\delta r < 10^{-3}$ \cite{LiteBIRD}. For $N_e = 60$ and $65$, the pivot-scale value obtained for all three regularizations in both approaches, $r_{\star} \simeq 4.3\times 10^{-3}$, is about four times this sensitivity. If the quoted uncertainty applies, a primordial $B$-mode signal of this amplitude is therefore within the reach of LiteBIRD. A detection would measure the amplitude of the primordial gravitational waves and would test the tensor sector of the inflationary dynamics driven by the Starobinsky potential in LQC models.

It is worth comparing these predictions with the present observational status of $r$. The bound $r < 0.036$ \cite{BICEPKeckData} is set by the recombination bump at $\ell \sim 80$, where the $B$-mode signal is strongest and the modes that exited the horizon during inflation are measured with the best sensitivity. By contrast, the regions of the spectrum where the three regularizations differ lie at $\ell \lesssim 20$ in the reionization bump, and they are governed by the modes in the upper part of the amplified oscillatory regime. The infrared plateau of mLQC-I lies at scales beyond the present horizon and does not appear in $C^{BB}_\ell$. The current $B$-mode bandpowers in the $\ell \lesssim 20$ window, shown as the black dots in Figs.~\ref{FIG:Cbb_Hybrid_60_FO+DS+FO}--\ref{FIG:Cbb_DressedMetric_60_FO+DS+FO} for the bandpower $\mathrm{BK}_{150}\times\mathrm{P}_{143}$, carry uncertainties far larger than the differences among the model curves. The models studied here therefore predict a signal that is large enough to be targeted, while concentrating their distinguishing features in the part of the spectrum that is presently inaccessible.

Even for upcoming experiments, discrimination among the models is unlikely to come from the $B$-mode power spectrum alone. At $\ell \lesssim 20$ the number of available sky modes is small, so cosmic variance sets a severe floor on any measurement, and extracting a primordial signal additionally requires delensing of the lensing $B$ modes and control of the Galactic dust and synchrotron foregrounds. While a detection of $r$ at the amplitude predicted here would be consistent with the framework studied in this manuscript, distinguishing LQC, mLQC-I and mLQC-II through their angular $B$-mode spectra is unlikely to be achievable with these experiments alone. In this sense, the tensor-to-scalar ratio at the pivot scale, which for $N_e = 60$ and $65$ does not depend on the regularization, is the observable most likely to be measured, whereas the model-dependent features survive only in the part of the spectrum that is hardest to access. Any future improvement in the sensitivity at low multipoles would improve the ability to test the amplified oscillatory part of the tensor spectrum. The infrared regime itself corresponds to scales far beyond the present horizon and is not directly accessible through the $B$-mode power spectrum.

Finally, the superhorizon tensor modes that carry the record of the bounce are not necessarily beyond observational reach altogether. In an expanding universe, a tensor mode that has crossed outside the horizon keeps its amplitude constant, so the enhancement or suppression acquired during the pre-inflationary phase is preserved rather than erased. Modes with wavelengths comparable to the present horizon are probed only through the lowest multipoles, where the small number of available sky modes limits the precision of any measurement of the power spectrum, while modes far beyond the present horizon are not accessible through the two-point function at all. The enhanced long-wavelength tensor modes computed here may, however, leave an imprint on higher-order correlation functions, such as a tensor bispectrum or a correlation between tensor and scalar modes. Such observables would be sensitive to the transient evolution of the effective mass near the bounce, where the three regularizations differ. To our knowledge, the corresponding computation for the tensor modes of mLQC-I and mLQC-II in both the hybrid and the dressed metric approaches has not been carried out, and it would be a natural extension of the present work. It would also show whether the infrared plateau of mLQC-I, whose amplitude is fixed by the emergent de Sitter phase, leaves a distinguishable signature in higher-order observables.

\section{Summary}

The goal of this work was to determine how the ambiguities in the quantization of the Hamiltonian constraint of loop quantum cosmology (LQC) affect the primordial tensor perturbations, and whether any of the resulting effects can be identified in the CMB $B$-mode spectrum. We studied standard LQC, and the modified models mLQC-I and mLQC-II. All three models resolve the big bang singularity with a nonsingular bounce, but they differ in the effective background near the bounce and in the contracting branch. The spacetime is quasi-de Sitter in mLQC-I and approaches the classical contracting solution in standard LQC and mLQC-II. The regularization therefore changes the background around which the perturbations propagate, and it also changes the effective mass of the perturbations. A second ambiguity arises when perturbations are quantized on the quantum background. The hybrid and dressed metric approaches give different effective masses for the perturbation modes. The two coincide in the classical regime but can differ substantially when the curvature is Planckian, close to the bounce. On top of these, the initial state of the modes in the contracting branch is not unique, and its choice can affect the infrared part of the spectrum, particularly in mLQC-I. For these reasons, the regularization of the constraint, the quantization of the perturbations, and the initial state need to be studied together instead of one at a time. The scalar perturbations of the three models have already been compared \cite{Li:2024lqc}. We studied the tensor perturbations for three reasons. First, they are determined by the same background but are governed by a different effective mass, so they test the effective dynamics independently of the scalar sector. Second, they are sensitive to the choice of regularization and of perturbation prescription in the same regime, so comparing hybrid and dressed metric results for all three models separates features that are robust from those that depend on a particular choice. Third, at linear order the primordial tensor modes are the only source of the primordial $B$ modes in the CMB, so the tensor power spectrum $\mathcal P_T(k)$, the tensor-to-scalar ratio $r(k)$, and the $B$-mode spectrum $C_\ell^{BB}$ connect the quantum geometry of the bounce to quantities that observations can in principle probe. To make this comparison meaningful, we used the same Starobinsky potential, matched background conventions, and a common pivot-scale normalization for all three models, and we followed the tensor modes from the high-curvature regime through the bounce into slow-roll inflation. Any difference among the results can then be attributed to the regularization, to the perturbation prescription, or to the initial state, and not to a difference in setup. We do not assume at the outset that these differences are large enough to be observed. Part of the aim was to find out how much of the model dependence survives in the observable spectra.

The three ambiguities enter the tensor equation in different places. The
regularization determines the effective background, the choice of perturbation
prescription determines the tensor effective mass, and the initial state sets the
boundary condition in the contracting branch. Standard LQC and mLQC-II approach
the classical regime in the contracting branch in qualitatively similar ways. The
mLQC-I model has an emergent quasi-de Sitter contracting branch, which changes the
asymptotic tensor frequency and with it the appropriate initial state. For LQC and
mLQC-II we use zeroth-, second-, and fourth-order adiabatic states, which measure
the initial-state dependence. For mLQC-I we use the de Sitter mode function in the
dressed metric approach and a Hankel-function solution in the hybrid approach. The
tensor effective mass connects the background to the spectrum. In the dressed
metric approach it is determined by $-a''/a$, and in the hybrid approach by the
combination $\rho-3P$ evaluated on the effective background of each model. The two
agree in the classical slow-roll regime and can differ substantially near the
bounce, where the curvature-induced mass is comparable to $k^2/a^2$. The infrared
tensor spectrum is therefore more sensitive to the bounce dynamics than the
scale-invariant regime, and it is the regime in which the effects of the three
ambiguities have to be separated.

The primordial spectra show which effects are robust and which are not. In the
scale-invariant regime, all models and approaches have the same amplitude
$A_T \simeq 9.02\times10^{-12}$, so the classical inflationary limit is recovered
independently of the regularization, the prescription, and the initial state. The
model dependence is at low wavenumber, where the tensor modes retain a record of
the bounce, the pre-inflationary branch, and the effective mass. The effect of the
regularization is seen for LQC and mLQC-II: for all adiabatic orders, mLQC-II has
larger amplitudes than LQC in the infrared and amplified oscillatory regimes. The
effect of the prescription is seen in the amplified oscillatory regime, where the
dressed metric spectra are larger than the hybrid ones. The effect of the initial
state is seen in the dependence of the infrared amplitude on the adiabatic order,
and in the isolated spikes of the second- and fourth-order states, which come from
the zeros of $W_k$. These spikes are properties of the state and not predictions
of the models. The mLQC-I model is the case in which the choice of prescription
and of state cannot be separated, because the two approaches use different initial
states. In the dressed metric approach, the de Sitter state gives an infrared
plateau whose amplitude is set by the Planckian de Sitter Hubble rate of the
contracting branch, in agreement with the analytic estimate (\ref{eq:plateau}).
This parallels the results in the scalar spectrum \cite{Li:2024lqc}. In the hybrid approach, the
Hankel-function state gives no plateau and a suppressed infrared spectrum. The
plateau is therefore a property of the dressed metric de Sitter state and not a
generic feature of mLQC-I. The stability analysis of Ref.~\cite{Pan:2025mlqci}
supports this state in the dressed metric approach, and the hybrid approach
remains to be studied.

The tensor-to-scalar ratio shows the same pattern. For $N_e = 60$ and $65$, the
pivot-scale ratio is $r_\star \simeq 4.3\times10^{-3}$ for all three models and
both approaches (Table \ref{TAB:r_star_list}), below the bound $r < 0.036$. It is
fixed in the scale-invariant regime, where all models have the same $A_s$ and
$A_T$, and we computed it for these two values of $N_e$ only. The scale
dependence of $r(k)$ carries more information than its pivot value. In the
scale-invariant regime, $r(k)$ grows slowly, as expected for slow-roll inflation,
and this part is the same for all models, approaches, and initial states. In the
upper part of the amplified oscillatory regime, $r(k)$ oscillates around $r_\star$
by a factor of a few. For LQC and mLQC-II this oscillation is the same for all
three adiabatic orders, so it is an imprint of the bounce that comes from the
different effective masses of the scalar and tensor modes. At smaller $k$, the
peak and the trough of $r(k)$ occur at wavenumbers that do not depend on the
adiabatic order or the approach, and their heights depend on the state. The
isolated spikes are properties of the state. For LQC and mLQC-II the hybrid and
dressed metric curves of $r(k)$ have nearly the same shape. For mLQC-I the two
approaches differ at small $k$, consistent with their different effective masses
and initial states in the contracting branch.

These differences are carried into the CMB through a common tensor transfer
function. For $N_e = 60$, the $B$-mode spectra of all three models match the
standard $\Lambda$CDM prediction at large multipoles and deviate from it at
$\ell \lesssim 20$. The mLQC-II model deviates most and mLQC-I least, in both
approaches, and the deviations are larger in the dressed metric approach. This
follows the ordering of the tensor amplitudes in the amplified oscillatory regime
and the ordering found for the scalar spectra \cite{Li:2024lqc}. The pivot-scale
value $r_\star$ is about four times the LiteBIRD target uncertainty
$\delta r < 10^{-3}$, so a primordial signal of this amplitude is within reach if
the target is met. Since $r_\star$ is the same for all three regularizations, such
a measurement tests the inflationary phase and not the regularization. The
features that distinguish the models lie at $\ell \lesssim 20$, where few sky
modes are available and a measurement requires delensing and foreground control.
Distinguishing the regularizations through the $B$-mode power spectrum alone is
therefore unlikely.

In summary, we find that the classical inflationary limit is recovered at high
wavenumber in all three models and both approaches, with the same pivot-scale
tensor-to-scalar ratio $r_\star \simeq 4.3\times10^{-3}$ for both $N_e=60$ and $65$,
independently of the regularization and of the perturbation prescription,  since for a fixed inflaton mass the normalization $\mathcal{P}_S(k_\star)=A_s$ fixes the number of e-folds between pivot exit and the end of inflation. 
 Long-wavelength
tensor modes retain a record of the bounce and, in mLQC-I, of the emergent de Sitter phase
before it. This record depends on the regularization, on the perturbation prescription,
and on the initial state. Part of it, namely the isolated spikes of the second and fourth
order adiabatic states and, in mLQC-I, the infrared plateau, is set by the state and is not
a prediction of the model. Only a small part of the model dependence is accessible through
$C_\ell^{BB}$, and it is confined to $\ell \lesssim 20$. During inflation a superhorizon
tensor mode keeps a constant amplitude, so the enhancement or suppression acquired near the
bounce is preserved. The infrared modes, however, lie far outside the present horizon and
cannot be probed through the power spectrum. They may leave an imprint on higher-order
correlation functions. In LQC the bounce has been shown to generate a scale dependent and
oscillatory scalar bispectrum, which is more sensitive to the bounce than the power
spectrum
\cite{Agullo:2015NG,Agullo:2018NG,Zhu:2018NG,Wu:2018NG,Agullo:2020fbw,Roshna:2023NG}.
To our knowledge, the tensor bispectrum and the mixed tensor--scalar correlations have not
been computed in LQC, and the scalar bispectrum has not been computed for mLQC-I and
mLQC-II. It will be interesting to compute these quantities for all three models in both
the hybrid and the dressed metric approaches, and to see whether the infrared plateau of
mLQC-I and the other model-dependent features leave a distinguishable signature in them.

\begin{acknowledgements}
We thank Bao-Fei Li, Meysam Motaharfar, Kreesha Panchal and Sahil Saini for useful discussions. This work is supported by NSF grant PHY-2409543.
\end{acknowledgements}

%\bibliographystyle{unsrt}
%\bibliography{lqc_tensor}

\end{document}